\documentclass{aa}
\usepackage{graphicx}

\newcommand \heI{\ion{He}{i}~}
\newcommand \heIn{\ion{He}{i}}
\newcommand \hal{H$\alpha$~}
\newcommand \haln{H$\alpha$}
\newcommand \hbeta{H$\beta$~}

\newcommand \kms{km s$^{-1}$~}
\newcommand \kmsn{km s$^{-1}$}
\newcommand \msun{M$_\odot$}
\newcommand \msunyr{M$_\odot$yr$^{-1}$}
\newcommand \rsun{R$_\odot$}
\begin{document}
\title{Eclipses by circumstellar material in the T Tauri star AA
  Tau. \\ II.  Evidence for non-stationary magnetospheric
  accretion\thanks{Based on observations obtained at Observatoire de
  Haute Provence (CNRS, France), Mt Maidanak Obs. (Uzbekistan), Calar
  Alto Obs. (Spain), Teide Obs. (Spain), Byurakan Obs. (Armenia),
  Assy-Turgen Obs. (Kazakstan), ESO La Silla (Chile), Lick Obs. (NOAO,
  USA), Tautenburg Obs. (Germany) and Roque de los Muchachos
  Obs. (Spain). }}

   \subtitle{ }

   \author{J. Bouvier\inst{1},
          K.N. Grankin\inst{2},
          S.H.P. Alencar\inst{3},
          C. Dougados\inst{1},
          M. Fern\'andez\inst{4},
          G. Basri\inst{5},
          C. Batalha\inst{6},
          E. Guenther\inst{7},
          M.A. Ibrahimov\inst{2},
          T.Y. Magakian\inst{8},
          S.Y. Melnikov\inst{2},
          P.P. Petrov\inst{9},
          M.V. Rud\inst{10}, 
          \and
          M.R. Zapatero Osorio\inst{11}
          }

   \offprints{J. Bouvier}

   \institute{Laboratoire d'Astrophysique, Observatoire de Grenoble, Universit\'e Joseph Fourier,
              B.P. 53, F-38041 Grenoble Cedex 9, France
         \and
             Astronomical Institute of the Academy of Sciences of Uzbekistan, Astronomicheskaya 33, 
             Tashkent 700052, Uzbekistan
         \and
             Departamento de F\'{\i}sica -- ICEx -- UFMG,
                 Caixa Postal 702, 30161-970, Belo Horizonte, Brasil
         \and
             Instituto de Astrof\'{\i}sica de Andaluc\'{\i}a, CSIC, Apdo. 3004, 18080 Granada, Spain
         \and
             Department of Astronomy, University of California at Berkeley, 601 Campbell Hall 
             3411, Berkeley, CA 94720, USA
         \and 
             Observat\'orio Nacional/CNPq, Rua General Jos\'e Cristino 77, Rio de Janeiro, RJ 
             20920-400, Brazil
         \and 
             Th\"uringer Landessternwarte Tautenburg, Karl-Schwarzschild-Observatorium, Sternwarte 5, 
             D-07778 Tautenburg, Germany
         \and
             Byurakan Astrophysical Observatory, Aragatsotn prov., 378433 Armenia
         \and 
             Crimean Astrophysical Observatory and Isaac Newton Institute of Chile, Crimean Branch, 
             p/o Nauchny, Crimea 98409, Ukraine
         \and 
             Fessenkov Astrophysical Institute, 480068 Almaty, Kazakstan
         \and
             LAEFF-INTA, P.O. Box 50727, E-28080 Madrid, Spain
             }
   \authorrunning{Bouvier et al.}
   \titlerunning{Magnetospheric accretion in AA Tau }

   \date{Received 2003; accepted 2003}
   
   \abstract{We report the results of a synoptic study of the photometric
     and spectroscopic variability of the classical T Tauri star AA Tau on
     timescales ranging from a few hours to several weeks.  The AA Tau
     light curve had been previously shown to vary with a 8.2d period,
     exhibiting a roughly constant brightness level, interrupted by
     quasi-cyclic fading episodes, which we interpreted as recurrent
     eclipses of the central star by the warped inner edge of its accretion
     disk (Bouvier et al. 1999). Our observations show the system is
     dynamic and presents non-stationary variability both in the photometry
     and spectroscopy. \\The star exhibits strong emission lines that show
     substantial variety and variability in their profile shapes and
     fluxes.  Emission lines such as \hal and \hbeta show both infall and
     outflow signatures and are well reproduced by magnetospheric accretion
     models with moderate mass accretion rates ($10^{-8}-10^{-9} {\rm
       M}_\odot {\rm yr}^{-1}$) and high inclinations ($i \ge 60\degr$).
     The veiling shows variations that indicate the presence of 2
     rotationally modulated hot spots corresponding to the two
     magnetosphere poles. It correlates well with the HeI line flux, with
     B-V and the V excess flux. We have indications of a time delay between
     the main emission lines (\haln, \hbeta and \heIn) and veiling, the
     lines formed farther away preceding the veiling changes. The time
     delay we measure is consistent with accreted material propagating
     downwards the accretion columns at free fall velocity from a distance
     of about 8R$_\star$. In addition, we report periodic radial
       velocity variations of the photospheric spectrum which might point
       to the existence of a 0.02\msun\  object orbiting the star at a
       distance of 0.08 AU. \\ During a few days, the eclipses
     disappeared, the variability of the system was strongly reduced and
     the line fluxes and veiling severely depressed. We argue that this
     episode of quiescence corresponds to the temporary disruption of the
     magnetic configuration at the disk inner edge. The smooth radial
     velocity variations of inflow and outflow diagnostics in the \hal
     profile yield further evidence for large scale variations of the
     magnetic configuration on a timescale of a month.  These results may
     provide {\it the first clear evidence for large scale instabilities
       developping in T Tauri magnetospheres as the magnetic field lines
       are twisted by differential rotation between the star and the inner
       disk.} The interaction between the inner accretion disk and the
     stellar magnetosphere thus appears to be a highly dynamical and time
     dependent process.
      }
   \maketitle
%

\section{Introduction}

T Tauri stars are low-mass stars with an age of a few million years at
most, still contracting down their Hayashi tracks towards the main
sequence. They are classified in two groups, the weak-line T Tauri stars
(WTTS) which merely exhibit enhanced solar-type activity and the classical
T Tauri stars (CTTS) which actively accrete material from a circumstellar
disk (see, e.g., M\'enard \& Bertout 1999). Understanding the accretion
process in young solar type stars, as well as the associated mass loss
phenomenon, is one of the major goals in the study of T Tauri stars.
Indeed, accretion has a significant and long lasting impact on the
evolution of low mass stars by providing both mass and angular momentum,
and the evolution and fate of circumstellar accretion disks around young
stars has become an increasingly important issue since the discovery of
extrasolar planets and planetary systems with unexpected properties.
Deriving the properties of young stellar systems, of their associated disks
and outflows is therefore an important step towards the establishment of
plausible scenarios for star and planet formation.


Early models assumed that the accretion disk of CTTS extended all the
way down to the star. 
However, the recognition that young stars have strong surface magnetic
fields of order of 1-3 kG (Johns Krull et al. 1999, 2001; Guenther et
al. 1999; Smith et al. 2003) raised the issue of the impact an
extended stellar magnetosphere might have on the structure of the
inner disk. 
%
Assuming that the main component of the stellar magnetosphere on the
large scale is a dipole, K\"onigl (1991) showed that, for typical mass
accretion rates in the disk (10$^{-9}$ to 10$^{-7}$\msunyr; Basri \&
Bertout 1989; Hartigan et al. 1995; Gullbring et al. 1998), the
magnetic torque exerted by the field lines onto the gaseous disk was
comparable to the viscous torque due to turbulence in the disk at a
few stellar radii (see also Camenzind 1990). Hence, the inner disk is
expected to be truncated by the magnetosphere at a distance of a few
stellar radii above the stellar surface. From there, material is
channelled onto the star along the magnetic field lines, thus giving
rise to magnetospheric accretion columns filled with hot plasma. As
the accreted material in the funnel flow eventually hits the stellar
surface at free fall velocity, strong accretion shocks develop near
the magnetic poles.

Observational support for these predictions of the magnetospheric accretion
scenario in CTTS has been accumulating over the last decade (see Bouvier,
Alencar \& Dougados 2003 for a recent review). Magnetospheric cavities with
inner radii in the range 3-8R$_\star$ are called for to account for the
near-IR properties of CTTS systems (Bertout et al. 1988; Meyer et al.
1997). Inverse P Cygni profiles observed in the Balmer and Pashen line
profiles of CTTS, with redshifted absorption components reaching velocities
of several hundred \kmsn, point to hot gas free falling onto the star from
a distance of a few stellar radii (Edwards et al. 1994, Folha \& Emerson
2001).  Magnetospheric accretion models have indeed been successful in
reproducing the main characteristics of the emission line profiles of some
CTTS, which suggest that at least part of the line emission arises in
accretion columns (Hartmann et al. 1994; Muzerolle et al. 2001). Finally,
hot spots covering about 1\% of the stellar surface are thought to be
responsible for the rotational modulation of CTTS luminosity (Bouvier \&
Bertout 1989; Vrba et al. 1993) and are identified with the accretion
shocks expected to develop near the magnetic poles in the magnetospheric
accretion scenario.

The observational evidence for magnetically channelled accretion in
CTTS has led to the development of steady-state axisymmetric MHD
models which describe the interaction between the inner disk and a
stellar dipole. These models provide a framework to understand the
physical connection between accretion and mass loss in CTTS, with the
open magnetic field lines threading the disk carrying away part of the
accretion flow while the remaining part is channelled onto the star
(e.g. Shu et al.  1994; Ferreira 1997).
Synoptic studies of a few CTTS systems have revealed correlated time
variability of the inflow and outflow diagnostics, both being modulated
on a rotation timescale. This has been interpreted as evidence for an
{\it inclined\/} stellar magnetosphere disrupting the inner disk
(Johns \& Basri 1995a; Petrov et al. 1996; Oliveira et al. 2000;
Petrov et al. 2001).

Magnetically mediated accretion in CTTS is presumably more complex and
possibly much more time variable than depicted by axisymmetric,
steady-state MHD models. Expanding upon earlier models (Aly \&
Kuijpers 1990; van Ballegooijen 1994; Lynden-Bell \& Boily 1994),
recent numerical simulations of the disk/magnetosphere interaction
suggest that the magnetic field lines that connect the star to the
disk can be substantially deformed by differential rotation on short
timescales. One class of models thus predict that differential
rotation between the footpoints of the field lines, one being anchored
into the star the other into the disk, leads to field line expansion,
opening and reconnection which eventually restores the initial
(dipolar) configuration (e.g. Goodson et al. 1997; Goodson \& Winglee
1999). This magnetospheric inflation process is thus expected to be
cyclic on a timescale of a few rotation periods and to be accompanied
by both episodic outflows during the opening of the magnetic structure
and time dependent accretion onto the star (Hayashi et al.  1996;
Romanova et al. 2002). Other models, however, suggest that under the
action of differential rotation the field lines drift radially
outwards in the disk leading to magnetic flux expulsion (Bardou \&
Heyvaerts 1996).  The response of the magnetic configuration to
differential rotation mainly depends upon magnetic diffusivity in the
disk, a free parameter of the models which is unfortunately poorly
constrained by current observations.

Due mostly to the lack of intense monitoring of CTTS on proper
timescales, the observational evidence for a time dependent
interaction between the inner disk and the stellar magnetosphere is at
present quite limited.  Episodic high velocity outbursts, possibly
connected with magnetospheric reconnection events predicted by recent
numerical simulations, have been reported for a few systems based on
the slowly varying velocity shift of blueshifted absorption components
of emission line profiles on a timescale of hours to days (Alencar et
al. 2001; Ardila et al. 2002). Possible evidence for magnetic field
lines being twisted by differential rotation between the star and the
disk has been reported for SU Aur by Oliveira et al. (2000). 
Another possible evidence for magnetic field lines being twisted by
differential rotation thus leading to quasi-periodic reconnection processes
has been reported for the embedded protostellar source YLW 15 based on the
observations of quasi-periodic X-ray flaring (Montmerle et al. 2000).

Since magnetically dominated accretion occurs on a scale of a few
stellar radii ($\leq$ 0.1 AU) which, at the distance of the nearest
star forming region cannot be resolved yet by current telescopes, one
of the most fruitful approach to probe the structure and evolution of
this compact region is to monitor the variations of the system over
several rotation timescales.
We therefore started synoptic campaigns on a number of CTTS a few
years ago. Results of previous campaigns have been reported by Chelli
et al.  (1999) for DF Tau and by Bouvier et al. (1999, hereafter B99)
for AA Tau.  The latter object proved to be ideally suited to probe
the inner few 0.01 AU of the system~: due to its high inclination
($i\simeq$75$\degr$, see B99), the line of sight to the star
intersects the region where the inner disk interacts with the stellar
magnetosphere.  The peculiar orientation of this otherwise typical
CTTS maximizes the variability induced by the modulation of the
magnetospheric structure and thus provides the strongest constraints
on the inner disk and the magnetospheric cavity. During the first
campaign (B99), multicolor photometry was obtained with no
simultaneous spectroscopy. This led to the discovery of recurrent
eclipses of the central object with a period of 8.2 days. We
attributed these eclipses to the keplerian rotation of a non
axisymmetric warp at the inner disk edge which periodically obscures
the line of sight to the star. We further proposed that the warped
inner disk edge directly resulted from the interaction of the disk
with an inclined magnetosphere, an expectation promptly confirmed by
Terquem \& Papaloizou (2000, see also Lai 1999).  While this first
campaign provided insight into the structure of the inner disk on a
scale of about 0.1~AU and constrained the large scale structure of AA
Tau's magnetosphere, the lack of simultaneous spectroscopy prevented
us from investigating the accretion columns connecting the inner disk
to the star.

We therefore organized a new campaign on AA Tau during the fall of
1999 combining simultaneous photometric and spectroscopic monitoring
over several rotation periods. One goal was to further investigate the
magnetospheric accretion region and relate the inner disk warp to
accretion columns and accretion shocks in a consistent way. Another
goal was to investigate the stability of the magnetospheric accretion
process on a month timescale, the duration of the campaign, as well as
on much longer timescales by comparing the results of the 2 campaigns
separated by 4 years.

The results of the campaign performed in 1999 are described in this paper.
Section 2 briefly describes the aquisition of photometric and spectroscopic
observations at various observatories, over a period of 5 months for
photometry and simultaneously over a period of 1 month for spectroscopy.
Section 3 presents the results of the spectroscopic and photometric
variability observed on timescales ranging from hours to months. Section 4
discuss the origin of the variability of the system and its relevance to
the magnetospheric accretion process. We argue that the main source of
photometric variability is variable circumstellar extinction which is
ascribed, as for the previous campaign, to the recurrent occultation of the
central star by the warped inner disk. The spectroscopic variability
provides evidence for magnetospheric accretion columns and associated hot
spots. In addition, we find that the accretion process is time dependent
and smoothly varies on a time scale of a month. We argue that the time
dependent accretion rate onto the star results from the development of
large scale instabilities in the magnetospheric structure, reminiscent of
the magnetospheric inflation cycles predicted by recent numerical
simulations. Section 5 concludes that the interaction between the inner disk
and the star's magnetosphere is a highy dynamical and time dependent
process and mentions a few implications of this result.

\section{Observations}

We describe in this section the multi-site campaign of
observations. Due to the numerous observatories involved and to the
variety of intrumentation we used, we provide only a brief account of
the data aquisition and reduction procedures at each site. The journal
of the observations is given in Table \ref{journal}.

   \begin{table*}
      \caption[]{Journal of observations.}
         \label{journal}
     $$        
         \begin{tabular}{llll|}
            \hline 
            \multicolumn{4}{c}{\bf Photometry} \cr
            \noalign{\smallskip}
            {Dates} & {Site} & {Telescope} & {Filters}\cr
            (1999) &&& \cr
            \noalign{\smallskip}
            \hline
            \noalign{\smallskip}
            09/08-16/12 & Mt Maidanak & 0.5m   & UBVR  \cr
            29/11-15/12  & OHP          & 1.20m     & UBVRI \cr
            25/11-02/12    & Calar Alto     & 1.50m          & BVRI  \cr
            26/11-04/01    & Teide      & 0.80m        & BVRI \cr
            30/11-08/12 & Byurakan   & 2.6m              & VRI \cr
            08/09-30/09  & Almaty     &  1m  &  BVR      \cr
            \noalign{\smallskip}
            \hline
            \end{tabular}
         \begin{tabular}{llll}
            \hline 
            \multicolumn{4}{c}{\bf Spectroscopy} \cr
            \noalign{\smallskip}
            {Dates} & {Site} & {Telescope} & {Spectral\ Range}\cr
            (1999) & & & (\AA) \cr
            \noalign{\smallskip}
            \hline
            \noalign{\smallskip}
            29/11-16/12        & OHP           & 1.93m            & 3890-6800  \cr
            25-28/11, 20-26/12 & ESO, La Silla & 1.52m        & 3530-9200  \cr
            26-27/11, 11-17/12 & Lick          & CAT              & 3850-10650 \cr
            28-29/11, 18/12 & Lick          & 3m               & 3850-10650 \cr
            27-28/11, 16-21/12    & Tautenburg    & 2m               & 4680-7100  \cr
            27-28/11    & Canarias      & NOT              & 4840-9400  \cr
            \noalign{\smallskip}
            \hline
         \end{tabular}
     $$
   \end{table*}
%


\subsection{Photometry}

The photometric observations were carried out over a period extending from
Aug. 9, 1999, to Jan. 4, 2000. AA Tau's light curve is best sampled in the
BVRI filters over the period from Nov. 25 to Dec. 15, 1999, when several
sites observed simultaneously (see Table \ref{journal}).

CCD photometry was performed at all sites except for Mt Maidanak and
Assy-Turgen observatories where a photomultiplier tube was used. After
images were suitably reduced (bias and flat-field corrected), differential
photometry between AA Tau and two reference stars was obtained using the
IRAF/DAOPHOT PSF fitting package.  The two reference stars are located less
than 2\arcmin\ away from AA Tau and recorded on the same images (these are
stars no.1125-01689518 and 1125-01691043 in the USNO2 catalogue). At Mt
Maidanak observatory, absolute photometry of AA Tau was obtained in the
UBVR filters, thus providing the required calibration of the differential
light curves derived from CCD photometry at other sites. The photometric
zero points of the differential UBVR light curves were thus derived from
measurements obtained simultaneously at Mt Maidanak and OHP. Since no
observations were performed in the I band at Mt Maidanak, the differential
I-band light curve cannot be calibrated this way. Instead, we assumed that
the average (V-I) color of AA Tau had not changed between 1995 (Bouvier et
al. 1999) and these new observations. This assumption is supported by the
fact that the average (V-R) color of AA Tau has remained the same between
the two epochs (see Section 3.1). The resulting calibrated light curves
have an rms photometric error of order of 0.02 mag in the BVR$_c$I$_c$
filters and up to 0.2 mag in the U-filter due to the system's faintness at
this wavelength. The I-band light curve might have a systematic photometric
error up to 0.05 mag due to the calibration method. The data are available
electronically at CDS Strasbourg.

Diaphragm photometry was performed at Maidanak Observatory
(Uzbekistan) from Aug.9 to Dec.16, 1999, using the 48cm telescope
equipped with a FEU-79 tube.  Measurements were obtained in the UBVR
bands with a diaphragm of 28\arcsec. Integration times ranged from 50
up to 120 seconds, depending on the filter. During one of the
photometric nights of the run, secondary standards were observed. The
data were reduced with standard procedures and assuming average
extinction coefficients for the site.  The final photometric error is
about 0.01 magnitude.

UBVRI observations were collected from Nov.29 to Dec.15, 1999, at
Observatoire de Haute-Provence (France) on the 1.2m telescope. The detector
was a 1k CCD camera yielding a field of view of 12\arcmin. The exposure
time ranged from 20 to 120 seconds depending on the filter and adapted to
seeing conditions in order to obtain a high signal to noise ratio on AA Tau
and the two comparison stars. CCD images were biased corrected and
flat-fielded with proper calibration images following a standard reduction
procedure.

Observations in the BVRI Johnson-Cousins system were carried out from Nov.
25 to Dec. 2, 1999 at the 1.5m telescope at the EOCA (Estaci\'on de
Observaci\'on de Calar Alto, Almer\'{\i}a, Spain) using a Tektronics
TK1024AB CCD, 1024x1024 pixels with a field of view 6.9x6.9\arcmin$^2$.
Integration times for the VRI filters were computed to maximize the S/N
ratio while remaining well within the linear region of the CCD.

Additional BVRI broad-band photometry of AA Tau was collected with the 0.8
m IAC80 telescope at Teide Observatory (Spain) on thirteen nights of
November through December 1999 and January 2000. We used the Thomson
detector (1024\,$\times$\,1024 pixel) mounted at the Cassegrain focus of
the telescope, which provides a pixel size of 0.43\arcsec~and a field of
view of 7.4x7.4\arcmin$^2$. Exposure times were typically 120--200\,s in
the BV bands, and 60--100\,s in the RI filters. Data were taken with a
seeing of 1.5--2.0\arcsec. We processed raw frames with usual techniques
within the IRAF environment, which included bias subtraction, flat-fielding
and correction for bad pixels by interpolation with values from the
nearest-neighbour pixels.

BVR observations of AA Tau were conducted with the 1.0 m "Carl Zeiss"
Jena telescope at the Assy-Turgen Observatory (2600m altitude) near
Almaty (Kazakstan) during four nights in September 1999. The single
channel photometer-polarimeter with the photocathode S20 operating in
photon counting mode was attached to the Cassegrain focus of the
telescope. Exposure times were 60s through a diaphragm of 20\arcsec.

Photometric observations in Byurakan (Armenia) were performed on 3 nights
between Nov.30 and Dec.8, 1999.  The 2.6m telescope was equipped with
the ByuFOSC-2 spectral camera, equipped with a 1060 x 514 CCD
(Movsessian et al. 2000) working in the imaging mode with a 12\arcmin
x 6\arcmin\ field of view.  Images were obtained in VRI filters and
were bias subtracted and flat-field calibrated following a standard
procedure.

\subsection{Spectroscopy}

The spectroscopic observations were carried out from Nov. 25 to
Dec. 26, 1999, from 5 sites. We obtained 54 high-resolution echelle
spectra of AA Tau over this period in quasi-simultaneity with the
photometry (see Table~\ref{journal}). Integration times ranged between
3600 and 5400s yielding an average S/N$\simeq$30 on the red continuum.

We obtained 23 spectra at the 1.93m OHP telescope with the ELODIE dual
fiber echelle spectrograph (Baranne et al. 1996) which yields 67 orders
covering the 3906 \AA\ to 6811 \AA\ domain at a mean spectral resolution of
$\lambda$/$\Delta \lambda$ $\simeq$ 42 000, and records simultaneously the
object and the neighboring sky.  Spectra were reduced with the automatic
on-line TACOS software (Queloz 1995). The reduction procedure includes
optimum extraction of the orders and flat-fielding achieved through a
Tungstene lamp exposure, wavelength calibration with a Thorium lamp
exposure, and removal of cosmic rays. Spectra are resampled every 0.03 \AA\ 
and corrected for the ELODIE transmission function. Cross-correlation
functions using the CORAVEL technique are also automatically computed.

Spectroscopic observations were also carried out at La Silla using the
1.52m ESO telescope with the FEROS spectrograph (Kaufer et al. 1998, 2000).
The mean resolution of the spectrograph is $\lambda / \Delta\lambda \approx
48000$, the spectral coverage is from 3500 \AA \, to 9200 \AA \, and the
exposure times varied from 45 to 60 minutes.  The reduction was
automatically performed on-line by the FEROS routines, which include
flatfielding, background subtraction, removal of cosmic rays, wavelength
calibration and barycentric correction.  Radial velocity corrections are
applied and all the data shown are in the stellar rest frame.  The spectra
are not flux calibrated, so each spectrum has been continuum normalized.

Some observations were carried out at Lick Observatory, using either
the 3m Shane reflector or the 0.6 m Coud\'e Auxiliary Telescope (CAT)
to feed the Hamilton Echelle Spectrograph (Vogt 1987) coupled
to a FORD $2048 \times 2048$ CCD. We recorded $\sim$92 orders covering
the optical spectrum from $\sim$3900 \AA\ to $\sim$8900 \AA.  The
mean resolution of the spectra is $\lambda / \Delta\lambda \approx
48,000$, and the exposure times varied from 45 to 75 min, depending on
the telescope used. The reduction was performed in a standard way
described by Valenti (1994) which includes flatfielding with
an incandescent lamp exposure, background subtraction, and cosmic ray
removal.  Wavelength calibration is made by observing a Thorium-Argon
comparison lamp and performing a 2-D solution to the position of the
Thorium lines as a function of order and column number.  Radial and
barycentric velocity corrections have been applied, and all the data
shown here are in the stellar rest frame.

Spectra were also obtained with the SOFIN echelle spectrograph (Tuominen
et al. 1999) at the 2.56m Nordic Optical Telescope (NOT).  The 3rd
camera was used, which provides a spectral resolution of about 12 \kms 
(R=26,000). The exposure time was 60 minutes. The useful spectral range
was from 4800 to 9400\AA\ with some gaps in the red.  The CCD images of
the echelle spectra were reduced with the 4A software package (Ilyin
2000). The standard procedure involves bias subtraction, correction
for the flat field, scattered light subtraction with the aid of 2-D
smoothing splines, elimination of cosmic spikes and correction for
vignetting function. The wavelength calibration was done with a Th-Ar
comparison spectrum.

Additional observation were carried out with the cross-dispersed Coud\'e
Echelle spectrograph of the 2-m-Alfred-Jensch telescope of the Th\"uringer
Landessternwarte Tautenburg on 6 nights between Nov.27 and Dec.21, 1999.  A
2.0\arcsec slit was used together with the 1kx1k Tektronix CCD yielding a
spectral resolution of about $\lambda/\Delta \lambda$ 35,000 over the
wavelength range from 4630 to 7370 \AA.  Standard IRAF routines were used
to flat-field and wavelength calibrate the spectra.

\section{Results}
\subsection{Photometry}


   \begin{figure*}
   \centering
   \includegraphics[width=\textwidth]{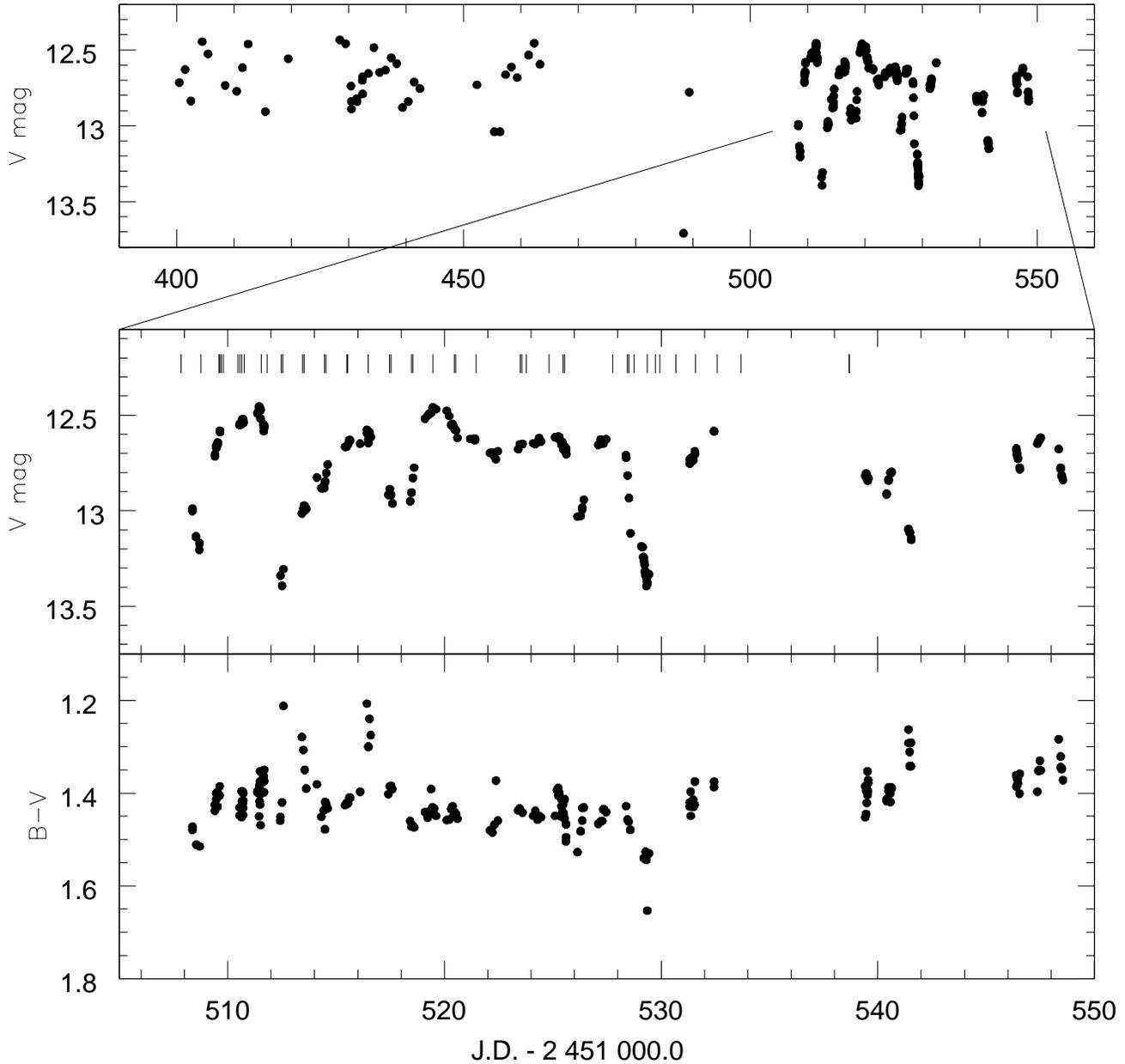}
      \caption{AA Tau light curve. The numbers on the x-axis  
        are the reduced Julian dates JD-2451000. Top: Full V-band
        light-curve. Middle: V-band light curve from JD 505 to 550; the
        vertical lines above the light curve indicate the dates of
        spectroscopic observations.  Bottom: B-V band light curve.}
         \label{lc}
   \end{figure*}
   
   AA Tau's V-band light curve recorded over 150 days is shown in
   Figure~\ref{lc} (top panel). During these 5 months, the maximum
   brightness level was roughly constant at V$\sim$12.4 and the photometric
   amplitude reached up to 1.0 magnitude. In the following, we mostly
   concentrate on the part of the light curve which has been the most
   heavily sampled by multi-site observations from J.D. 2451508 to 549,
   which is also when simultaneous spectroscopic observations were
   obtained.
   
   An enlargement of this section of the light curve is shown in
   Figure~\ref{lc} (middle panel). The photometric variations are
   nearly continously sampled over more than 3 weeks (from JD 508 to
   533) and reveal large-scale brightness fluctuations occurring on a
   time scale of a few days with a maximum amplitude of 1.0
   magnitude. This light curve bears some resemblance with the one we
   obtained in 1995 from a previous multi-site campaign (see Figure~1
   in Bouvier et al. 1999, hereafter B99).  The maximum brightness
   level in the V-band is the same as in 1995 and is interrupted by
   luminosity dips lasting several days.  Qualitatively similar dips
   were observed in 1995 with, however, a larger photometric amplitude
   (1.6 mag in 1995 compared to 1.0 mag in 1999).
   
   Another similarity between the 1995 and 1999 light curves is the (B-V)
   color which exhibits little change as the system's brightness varies,
   except for a few transient flaring-like episodes. This is shown in
   Figure~\ref{lc} (bottom panel) where the (B-V) light curve is seen to
   remain essentially flat within about 0.1 magnitudes while the star's
   brightness varies by more than one magnitude in the V-band. A few
   short-term episodes with an amplitude of about 0.2 magnitudes appear on
   J.D. 512/513 and 516 where the system suddenly turns bluer and on J.D.
   529 when it turns redder on a timescale of a few hours. The small
   amplitude (B-V) light curve which contrasts with the large luminosity
   variations had already been reported in the 1995 light curve, as well as
   the occurrence of transient blueing episodes (see B99, Figure~2).
   
\begin{figure}
     \centering \includegraphics[width=8cm]{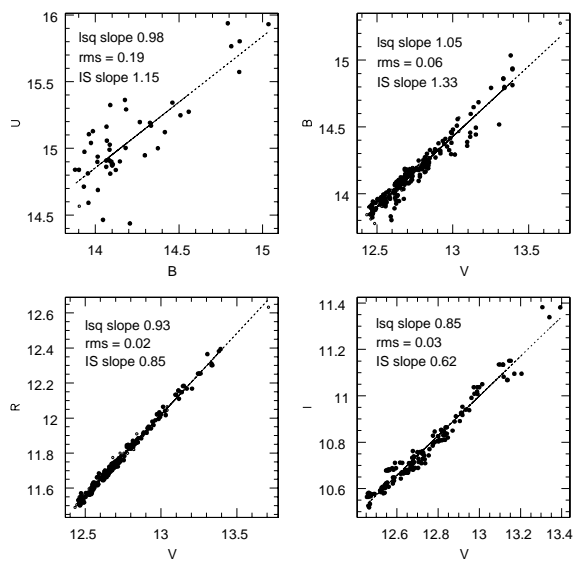}
      \caption{Correlated brightness variations in UBVRI filters. Small
        open dots: JD$\leq$500, large filled dots: JD 505-550. Least-square
        fits to the observed correlations are shown as a solid line. The
        slope and rms of the fit are given in each panel, as well as the
        expected slope for an interstellar reddening law (``IS slope'',
        Savage \& Mathis 1979).}
         \label{magslop}
   \end{figure}

   The photometric variations in the B, R, and I filters are similar
   to those observed in the V-filter. Figure~\ref{magslop} illustrates
   the observed correlation between the photometric variations in the
   various filters. The U-band measurements are affected by large
   photometric errors (of order of 0.2 mag due to the system's
   faintness at this wavelength) and will not be considered further.
   The slope of the linear least square fit to the observed
   correlations is given in each panel of Figure~\ref{magslop}
   together with the fit rms, and the slope expected from interstellar
   reddening (``IS slope'') is indicated.  The correlation is
   particularly tight for the V and R filters.  The B vs V diagram
   exhibits a well-defined upper envelope with some points
   ``dropping'' from this envelope, which corresponds to the blueing
   episodes mentionned above.  A noticeable feature of the I vs V
   diagram is the apparent change of slope at V$\sim$12.8.
   
\begin{figure}
   \centering
   \includegraphics[width=8cm]{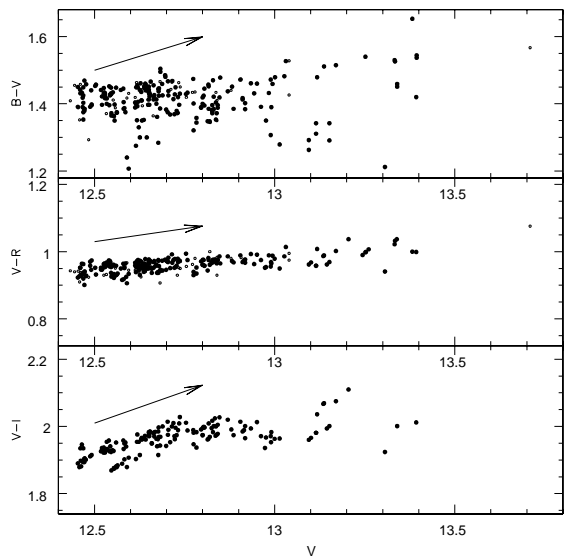}
      \caption{Color-magnitude diagrams. The amplitude of the vertical scale is the same in
        each panel. Reddening vectors are illustrated for A$_V$=0.3 mag.
        Small open dots: JD$\leq$500, large filled dots: JD 505-550.  }
         \label{cmds}
   \end{figure}

   A complementary representation of the color variations of the system is
   given in Figure~\ref{cmds} where the (B-V), (V-R) and (V-I) colors are
   plotted against V magnitude. With the exception of the blueing episodes,
   the near constancy within 0.1 mag of the (B-V) color is recovered in the
   (V, B-V) diagram.  The upper envelope of the points in this diagram
   suggests a possible trend for the system to become slightly redder at
   the lowest brightness levels, though with a much lower reddening slope
   than expected from insterstellar extinction. The average (B-V) color of
   the system is $\sim$1.42 in the 1999's light curve, while it was
   $\sim$1.25 in 1995, and the average (V-R) color has remained the same
   between the two epochs. This probably indicates a slightly lower
   accretion rate in 1999 than in 1995, thus decreasing the blue excess and
   yielding a slightly redder (B-V) color without afecting much (V-R).

   The (V, V-R) and (V, V-I) diagrams indicate that the system gets redder
   when fainter up to V$\sim$12.8, with a slope similar to that expected
   for interstellar extinction. Past this point, however, as the system's
   brightness further decreases the colors appear to saturate at a nearly
   constant value. An interesting feature of the (V, V-I) diagram is the
   evidence for two parallel tracks around V$\sim$12.6, where the system
   appears to oscillate between two (V-I) color states. This phenomenon
   occurs within a few hours on J.D. 511 at the start of a large luminosity
   dip and will be discussed further below.
   
   Overall, the luminosity and color variations observed in AA Tau in 1999
   are qualitatively similar to those reported by B99 for the 1995 light
   curve.  This leads us to believe that the dominant sources of
   photometric variability have not changed between the two epochs. A major
   difference, however, is that in 1995 the photometric variations were
   quasi-periodic on a timescale of $\sim$8.3-8.6 days. A periodogram
   analysis of the 1999's light curve reveals no significant period.  Both
   the periodogram analysis (Scargle 1982) and the string-length method
   (Dworetsky 1983) suggest a best period of order of 16.5 days, but with a
   low confidence level.  Photometric periods previously reported for AA
   Tau range from 8.2 days (Vrba et al. 1989; Shevchenko et al. 1991) to
   8.3-8.6 days (B99). A periodogram analysis of AA Tau V-band light curve
   over 14 years, from 1987 to 2000, built from Mt Maidanak data also
   suggests a period of 8.1867 days (Grankin, priv.  comm.). Additional
   support to the existence a 8.2d period in the system is reported in the
   next section where we show that the radial velocity of the star appears
   to smoothly vary over this period.
   
   \begin{figure}
   \centering
   \includegraphics[width=8cm]{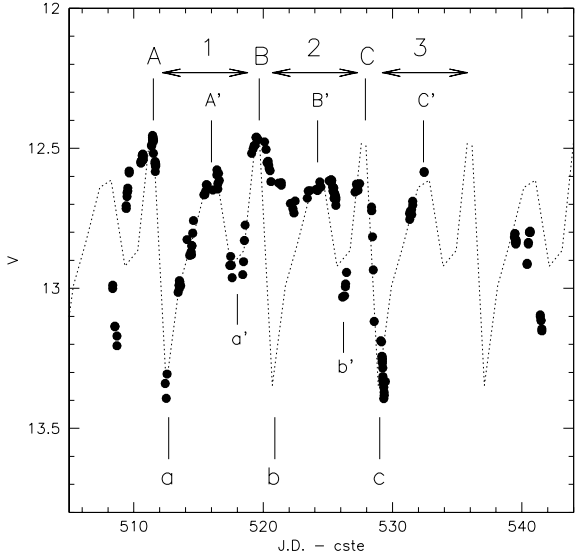}
      \caption{AA Tau's V-band light curve. Part of the light curve between
        JD 511.5 and 519.5 (``cycle 1'') has been replicated assuming a
        period of 8.2 days (dotted line). The various brightness minima and
        maxima are labelled with letters (see text).}
         \label{replicate}
   \end{figure}

   Figure~\ref{replicate} shows AA Tau's V-band light curve with an
   illustration of the expected photometric variations {\it
   assuming\/} a 8.2d period. The variations observed from J.D. 511 to
   519 have been replicated on the rest of the light curve assuming a
   8.2d period.  Several brightness maxima and minima appear to occur
   repeatedly on this timescale. This is the case for two main maxima
   (labelled A, B) and possibly a third one (C) at JD$\sim$511, 519,
   and 528 as well as for two secondary maxima (A', B') and possibly a
   third one (C') on JD$\sim$ 516, 524, and 532. The two deepest
   minima (a, c) are separated by $\sim$17 days, i.e., about twice the
   assumed period, and two secondary minima (a', b') are in phase with
   the 8.2d period.  Clearly, the phase coherence is lost prior to JD
   511 and after JD 535.  Hence, even though the light curve is not
   periodic, we do find evidence for a characteristic timescale for
   the photometric variations which is consistent with the period
   reported previously by several authors.
   
   The major discrepancy between the expected 8.2d period and the
   observed light curve is the absence of a deep minimum around JD 521
   (labelled ``b'' in Figure~\ref{replicate}). The flux of the system
   does decrease around this phase but only by about 0.2 magnitudes,
   i.e., much less than during the two deep minima observed 8.2 days
   before and after this date, respectively. In the following, we will
   refer to this part of the light curve between JD=519 and JD=525 as
   the ``photometric plateau''. Even more intringuing is the similar
   shape and depth of the two deep minima on JD 512 and 529, both
   being asymmetric with a rapid flux decrease and a more gradual
   return to maximum brightness. Assuming that the 8.2d period is
   intrinsic to the system and probably reflects the rotational period
   of the star, the light curve suggests that one of the major sources
   of photometric variability has disappeared for one cycle but was
   restored on the next one. We show in the next section that a
   similar conclusion is reached from the analysis of the
   spectroscopic data.

\subsection{Spectroscopy}


   \begin{figure*}
   \centering
   \includegraphics[width=14cm]{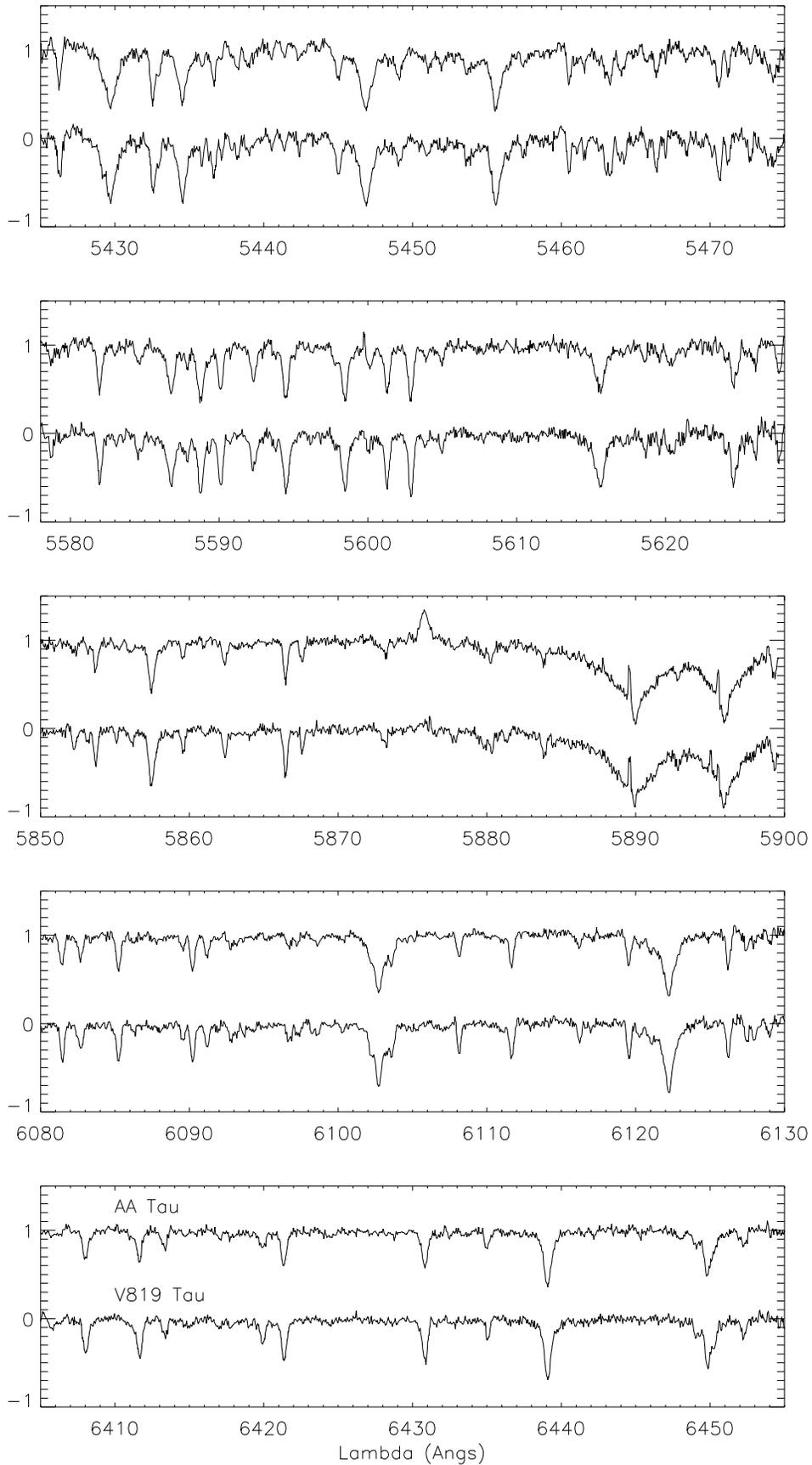}
      \caption{Samples of an ELODIE spectrum of AA Tau (upper curve) obtained on JD
        519.47. The spectrum of V819 Tau (lower curve), a K7 spectral
        template, is shown for comparison. In each spectral order the
        continuum has been normalized to unity and V819 Tau's spectrum has
        been shifted for clarity.  The spectral orders shown in this figure
        were used to derive veiling (see text). The HeI emission line
        ($\lambda$5876) appears in emission in AA Tau's spectrum.}
         \label{aa_phot_spec}
   \end{figure*}

AA Tau has been previously classified as a K7 dwarf (Kenyon \& Hartmann
1995). Its spectrum is that of a moderately active classical T Tauri star
which exhibits clear photospheric lines and a few major emission lines,
e.g. EW(\hal)$\simeq$10-20 \AA. The Balmer lines are characterized by the
presence of a deep central absorption feature in the emission profiles
(e.g. Edwards et al. 1994).

We discuss below the analysis of the 54 high resolution spectra of AA Tau
obtained during the campaign. The projected rotational velocity of the star
was measured from a correlation analysis of the photospheric spectrum.
Veiling was measured on the 33 ELODIE and FEROS spectra from the comparison
with a K7 template. The S/N ratio of all spectra was high enough to
investigate both the shape and the flux variations of the emission line
profiles, most notably \hal, \hbeta, and \heI.

\subsubsection{Photospheric Lines}

   \begin{figure}
   \centering
   \includegraphics{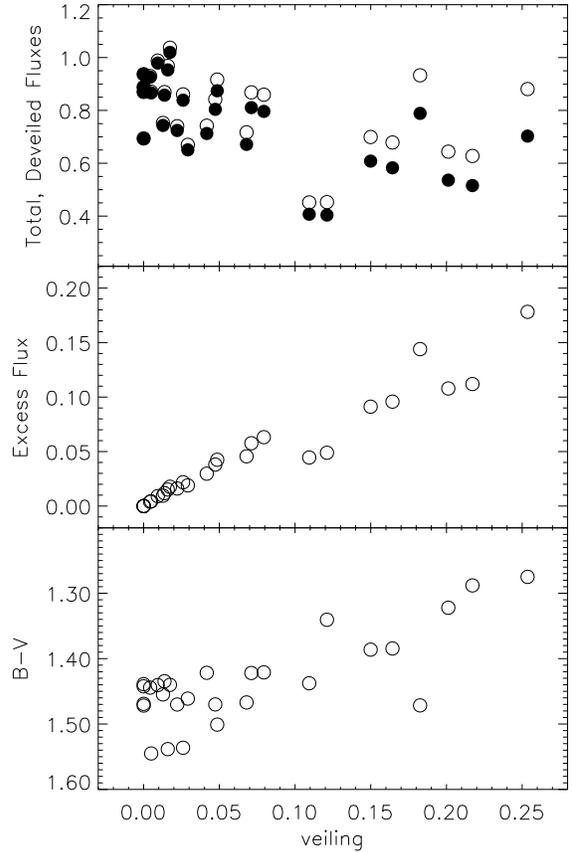}
      \caption{Total flux in V-band (top panel open circles),
        de-veiled flux (top panel filled circles), excess flux (middle
        panel) and B-V (lower panel) vs. veiling. 
The fluxes were
        interpolated on the veiling dates and are presented in
        arbitrary units. }
         \label{ex_veil}
   \end{figure}
%
   
   \begin{figure}
   \centering
   \includegraphics{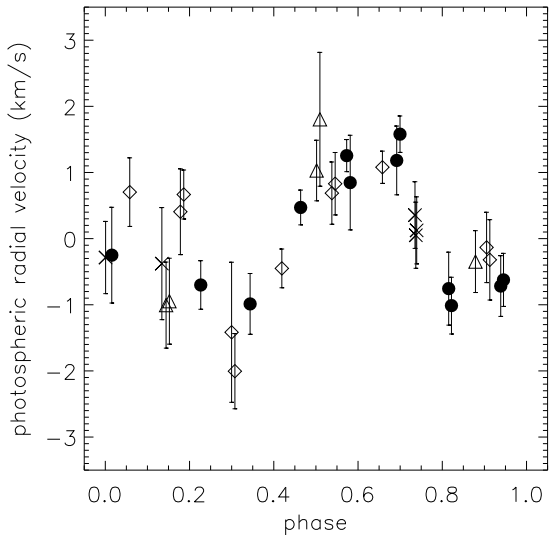}
      \caption{Photospheric radial velocities folded in phase with
        a 8.288 day period. Different symbols represent successive cycles.
        }
         \label{vrad_phot}
   \end{figure}
%
   
   We first derive estimates of the AA~Tau projected rotational velocity
   from 20 OHP spectra using the cross-correlation functions automatically
   computed with the ELODIE spectrograph and the calibration relation given
   in Queloz et al. (1998) for slow rotators. The derived {v$\sin$i} values
   range from 10.09 to 12.34 km s$^{-1}$ with an average of 11.3 km
   s$^{-1}$ (rms: 0.7 km s$^{-1}$).  This value fully agrees with the
   {v$\sin$i} of 11.4 km s$^{-1}$ computed by Hartmann \& Stauffer (1989).
   
   We compute the veiling, defined as the ratio of continuum excess flux
   over photospheric flux, using a $\chi^2$ fit method on five spectral
   intervals typically 50 \AA-wide\, located between 5500 and 6500 \AA\,
   and centered at: 5450, 5600, 5860, 6110 and 6420 \AA\ (see
   Fig.~\ref{aa_phot_spec}). We exclude from the fit the strong
   photospheric lines of NaD and CaI 6122\AA\, which Stout-Batalha et al.
   (2000) have shown could be enhanced by the accretion process. We perform
   veiling calculation on the 33 FEROS and ELODIE spectra alone. Both have
   comparable high spectral resolution, sufficient signal to noise (S/N) on
   the continuum level and sample the whole duration of the spectroscopic
   observations.  For each data set, we compute the veiling relative to a
   reference AA~Tau spectrum observed in the same instrumental
   configuration (an average of three spectra observed on JD 519.47,
   520.45, 520.52). The absolute veiling of this reference spectrum is then
   calibrated using the template weak line T~Tauri star V819~Tau observed
   with ELODIE.  V819~Tau (SpT=K7V) appears to be a very good spectral
   match to AA~Tau (Fig.~\ref{aa_phot_spec}).  Uncertainties associated
   with the relative variations of veiling, estimated from the rms of the
   five individual measurements, range from 0.01 to 0.05.  An additional
   uncertainty of typically 0.1 is however present on the absolute level of
   veiling.

The derived veiling values are low, ranging from undetectable to 0.3,
and do not correlate with the photometry (see Figure \ref{ex_veil} top
panel, open circles). The main source of photometric variations is
therefore not related to the continuum excess flux.  Two main
increases in veiling, lasting typically 3-4 days, occur at JD=513.5
and JD=516.5 during respectively the egress phase of the first large
photometric dip and just before the following small photometric dip
(see Fig.~\ref{line_flux}). The veiling also increases at the very
beginning and towards the end of the sampled light curve. During the
photometric ``plateau'' (JD 519-525) the veiling is extremely weak. We
observe a strong correlation of veiling with (B-V) color and HeI line
flux (see Figs. \ref{ex_veil} and \ref{tdelay_veil}).  We also combine
the measured veiling values with the V-band light curve to derive both
the underlying ``de-veiled'' photospheric flux and the flux of the
continuum excess in the V band (Fig.~\ref{ex_veil}). We included in
Figure \ref{ex_veil} only points where the photometry could be safely
interpolated at the time of the spectroscopic observations. The excess
flux variation closely follows the veiling one, with some scatter at
high veiling values which will be discussed below.

   We also compute, in the same wavelength intervals used to calculated the
   veiling, the radial velocity of photospheric lines by cross-correlation
   with the template spectrum, with a typical accuracy of 500 m s$^{-1}$
   (both our observing procedure and the correlation algorithm were not
   optimized for precise radial velocity measurements).
   
   From the combined ELODIE and FEROS spectra (33), we measure an average
   heliocentric radial velocity of 17.1 km s$^{-1}$ (rms of 0.9 km
   s$^{-1}$) consistent with the previous derivation of 16.1 $\pm$ 2.1 km
   s$^{-1}$ (Hartmann et al. 1986). The amplitude of variation of the
   photospheric radial velocity is small ($\sim$2 km s$^{-1}$) but
   significantly larger than the estimated measurement errors ($\simeq$0.5
   km s$^{-1}$). A string length analysis of the photospheric radial
   velocity variations yielded a most likely period of 8.288 days which is
   in agreement with the long term variability period found with the
   photometry. Figure \ref{vrad_phot} displays the photospheric radial
   velocity curve folded in phase with this period

\subsubsection{Survey of emisson line profiles}

\begin{figure*}
   \centering
   \includegraphics{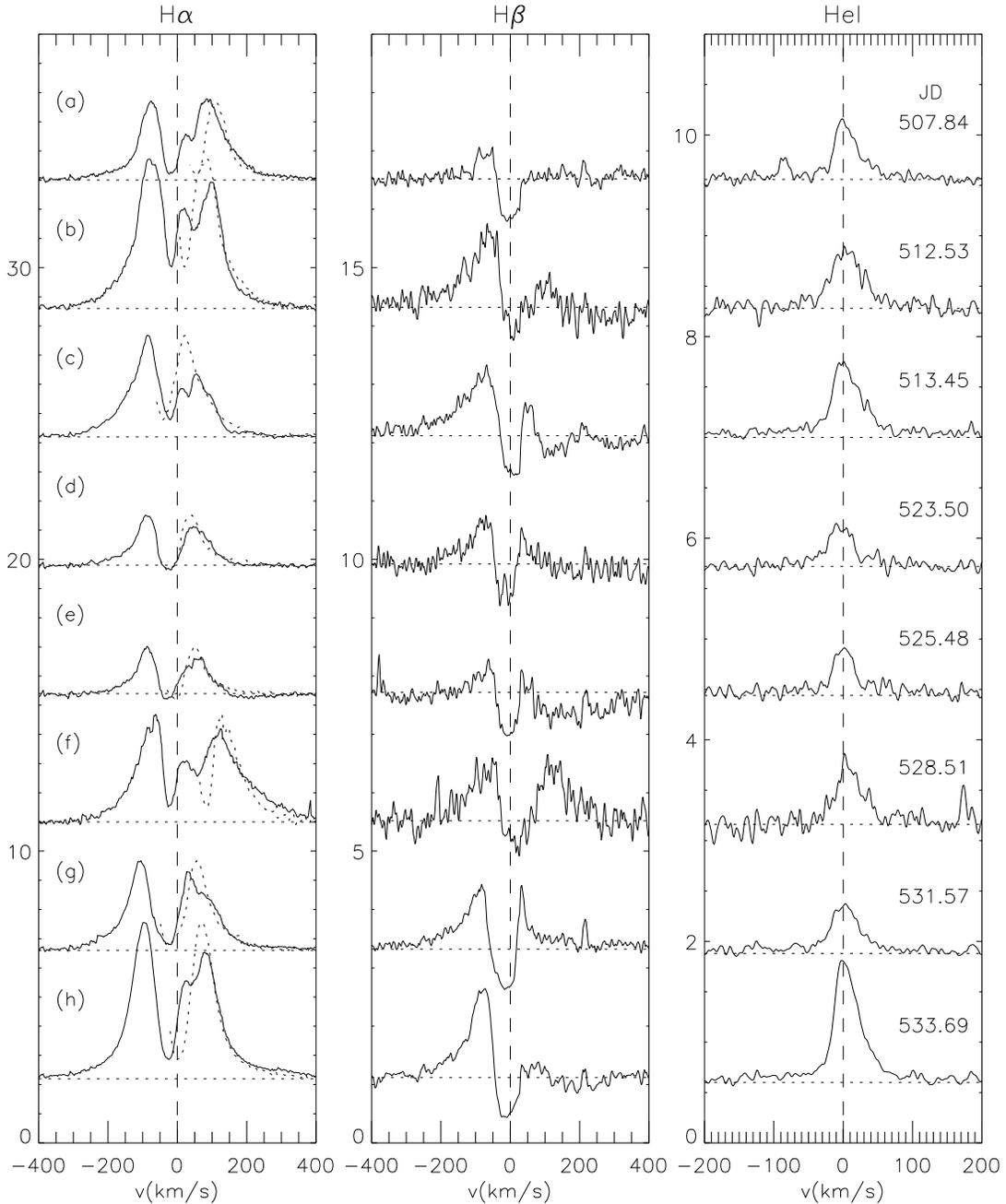}
      \caption{Sample of residual line profiles. The profiles have been
        shifted for clarity. The vertical dashed lines are the spectral
        line center at the stellar rest frame and the horizontal dotted
        lines show the continuum level. The \hal blue wing has been
        reflected with respect to line center over the red wing. The
        numbers on the right are the dates of observation JD-2451000.0.  }
         \label{profiles}
   \end{figure*}

A sample of the various residual line profiles discussed in the following 
sections is shown in Figure \ref{profiles}. Residual profiles were obtained
by continuum normalizing the AA Tau spectra and subtracting the normalized
and veiled K7 template star (V819 Tau) used in the veiling 
measurements. We do not have veiling measurements for all spectra due 
to the low S/N of some of the observations and in order to compute the 
residual profiles we used the nearest available veiling value. Since the 
observations are taken very close to each other and the highest measured 
veiling value is only about 0.3, we should not be using unreasonable 
values. We note in Figure \ref{profiles} that the emission line
profiles varied both in intensity and shape during our observations. 

%

   \begin{figure*}
   \centering
   \includegraphics{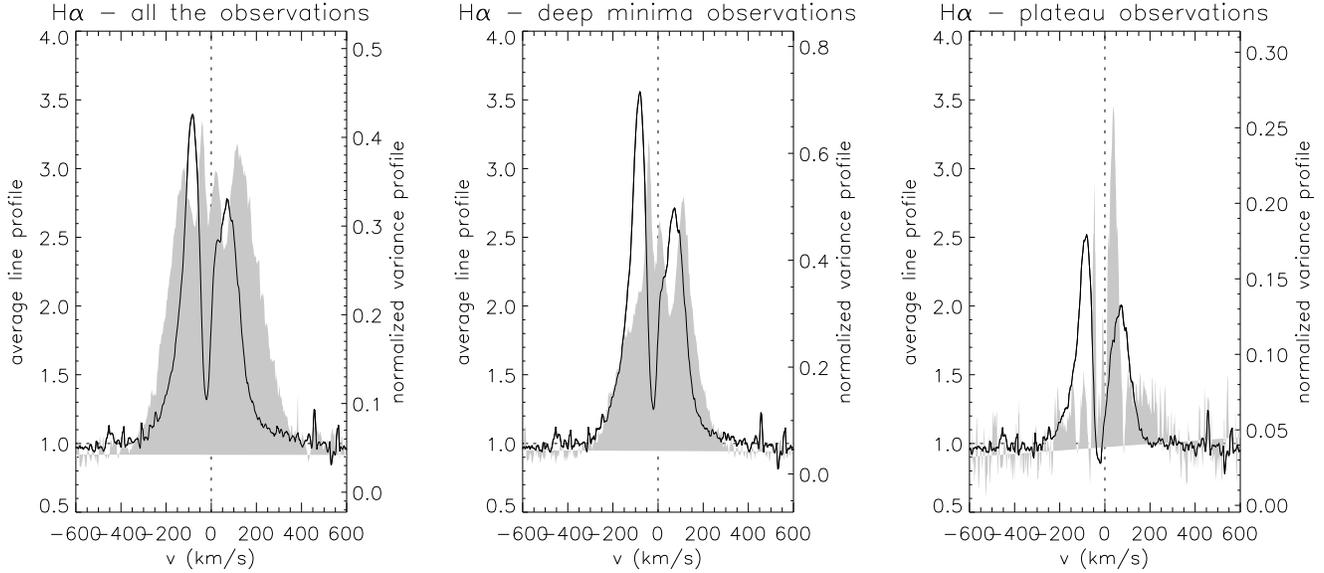}
      \caption{Average \hal line profiles (solid lines) and variance 
      profiles (grey shaded areas) calculated with residual spectra.
      Left: all the \hal observations. Middle: the \hal observations
      taken during the deep photometric minima. Right: the observations 
      taken during the photometric ``plateau'' (from JD=518 to JD=526).
              }
         \label{variance}
   \end{figure*}
%
   
   The \hal line presents double-peaked emission profiles, the blue
   emission peak generally more intense than the red one, resembling
   previously reported AA Tau profiles by Edwards et al. (1994),
   Muzerolle et al. (1998) and Alencar \& Basri
   (2000). The \hal profiles display both blueshifted and
   redshifted absorptions at low velocities in most of the observations as
   can be seen in Figure \ref{profiles}.
   The blueshifted absorption is more intense than the redshifted
   absorption component and is also always present, while the
   redshifted absorption once totally disappeared from our
   observations for 2 days (Fig. \ref{profiles}$d$). The outer wings
   of the \hal profiles tend to be symmetric, but some profiles do
   present asymmetries in the outer red wing either as a lack of
   emission (most commonly, Fig.  \ref{profiles}$c,e$) or as an extra
   red emission (Fig.  \ref{profiles}$f$).  The \hbeta line profiles
   are also double-peaked but display most of the time a single
   absorption component that is centered or slightly blueshifted.  A
   few \hbeta profiles present an extra redshifted absorption at high
   velocities (Fig. \ref{profiles}$c$,$e$) and these tend to
   correspond to the asymmetric \hal spectra that show a lack of
   emission in the outer red wing. The \heI line shows only a narrow
   component (NC) that is slightly redshifted. It is clearly
   asymmetric in some observations (Fig. \ref{profiles}$a$,$c$,$h$),
   with more emission in the red than in the blue side of the
   profile. Edwards et al. (1994) present AA Tau profiles of
   \heI with broad and narrow components but at a much higher veiling
   value (r=0.6) than we observed. The NaD lines (see
   Fig.~\ref{aa_phot_spec}) match very well the template spectrum and
   are basically photospheric.

\subsubsection{Variance profiles}

We show in Figure \ref{variance} the average residual line profiles of \hal
calculated with all the observations ($left$) and at different photometric
epochs~: during the deep minima ($middle$) and the photometric ``plateau''
($right$).  Also shown in the shaded area are the normalized variance
profiles as defined by Johns \& Basri (1995b), which measure the
amount of variability of each velocity bin in the line. The total \hal
profile shows a rather uniform variability, the red wing presenting only a
slightly more extended variability than the blue wing. The profiles taken
during the photometric ``plateau'' and the deep minima are quite different
from each other. The largest spectroscopic variations occur during the deep
photometric minima. In contrast, the line profile does not vary much during
the photometric ``plateau'' and its intensity is also much weaker. In
agreement with the photometry, this behavior indicates that from JD=518 to
526 a major source of variability of the system disappeared.

\subsubsection{Line Fluxes and Profile Decomposition}
   \begin{figure*}
     \centering \includegraphics{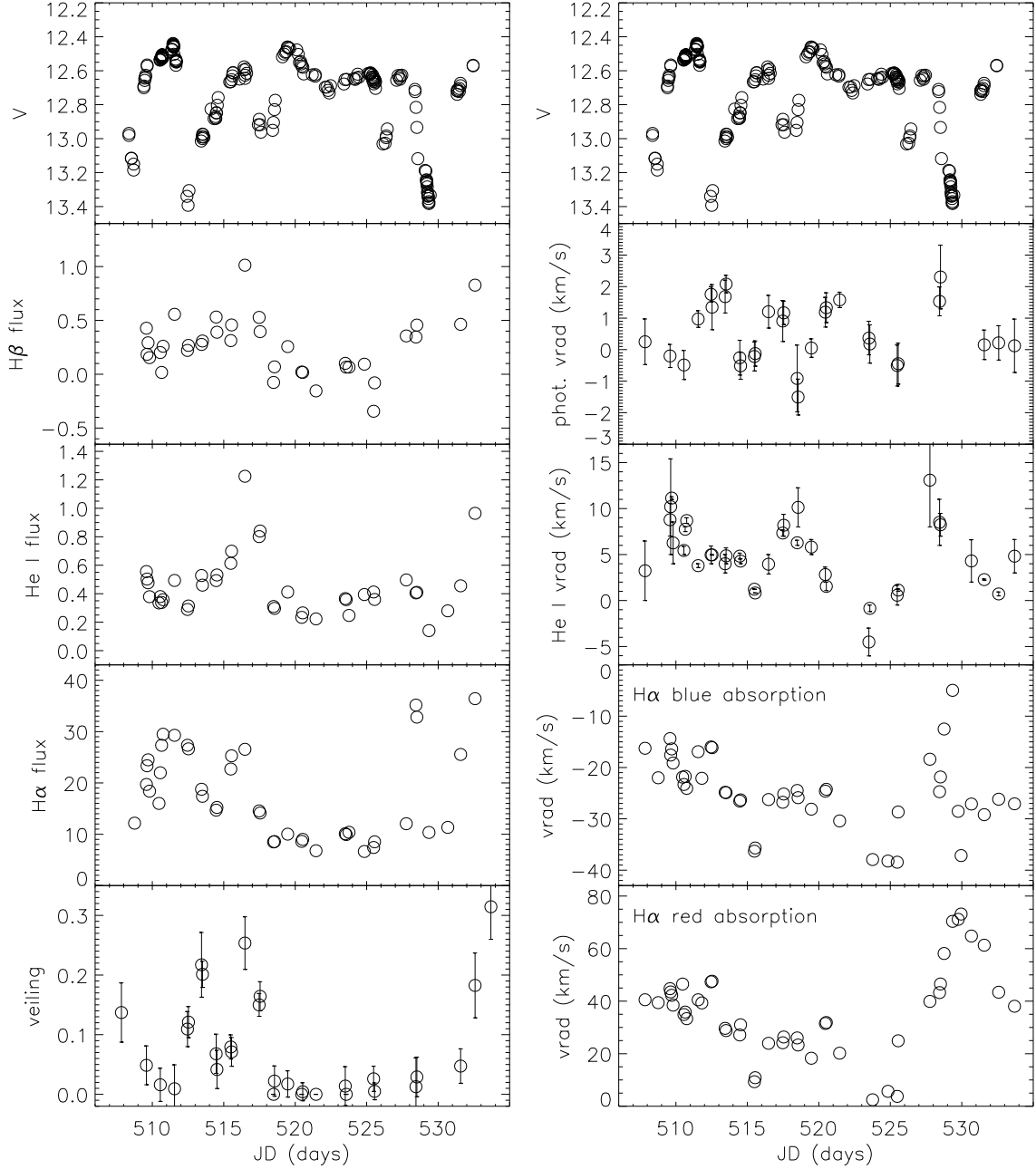}
      \caption{V magnitude, \haln, \hbeta, \heI line fluxes and veiling 
        (left panel) and V magnitude, photospheric radial velocity, \heI
        radial velocity, and radial velocity of the blue and red absorption
        components in the \hal profile (right panel). The line fluxes are
        in arbitrary units.}
         \label{line_flux}
   \end{figure*}
%

   We measured the total equivalent widths of the \haln, \hbeta and \heI
   lines in order to compute line fluxes with the photometric measurements
   as $F(H\alpha)=cst \times EW(H\alpha) \times 10^{-0.4m_R}$, $F(HeI)=cst
   \times EW(HeI) \times 10^{-0.4m_V}$ and $F(H\beta)=cst \times EW(H\beta)
   \times 10^{-0.4m_B}$, where $cst$ is an arbitrary constant and $m_R$,
   $m_V$ and $m_B$ are the R, V and B-band magnitudes of the system,
   respectively.  The photometric and spectroscopic measurements were not
   always simultaneous so we interpolated the light curves at the time of
   the spectroscopic observations in order to get the correct values for
   the magnitudes. No extrapolation was made.  The line fluxes obtained are
   presented in Figure \ref{line_flux}. We looked for periodicities in the
   line flux variations using the Scargle (1982) periodogram
   estimator as modified by Horne \& Baliunas (1986) that is
   appropriate to handle irregularly spaced data and the string-length
   method (Dworetsky 1983). Both methods yielded similar results,
   showing best period detections around 16 or 8.5 days but with rather
   high false alarm probabilities (a few percent) of being created by
   chance.
   
\begin{figure*}
   \centering
   \includegraphics{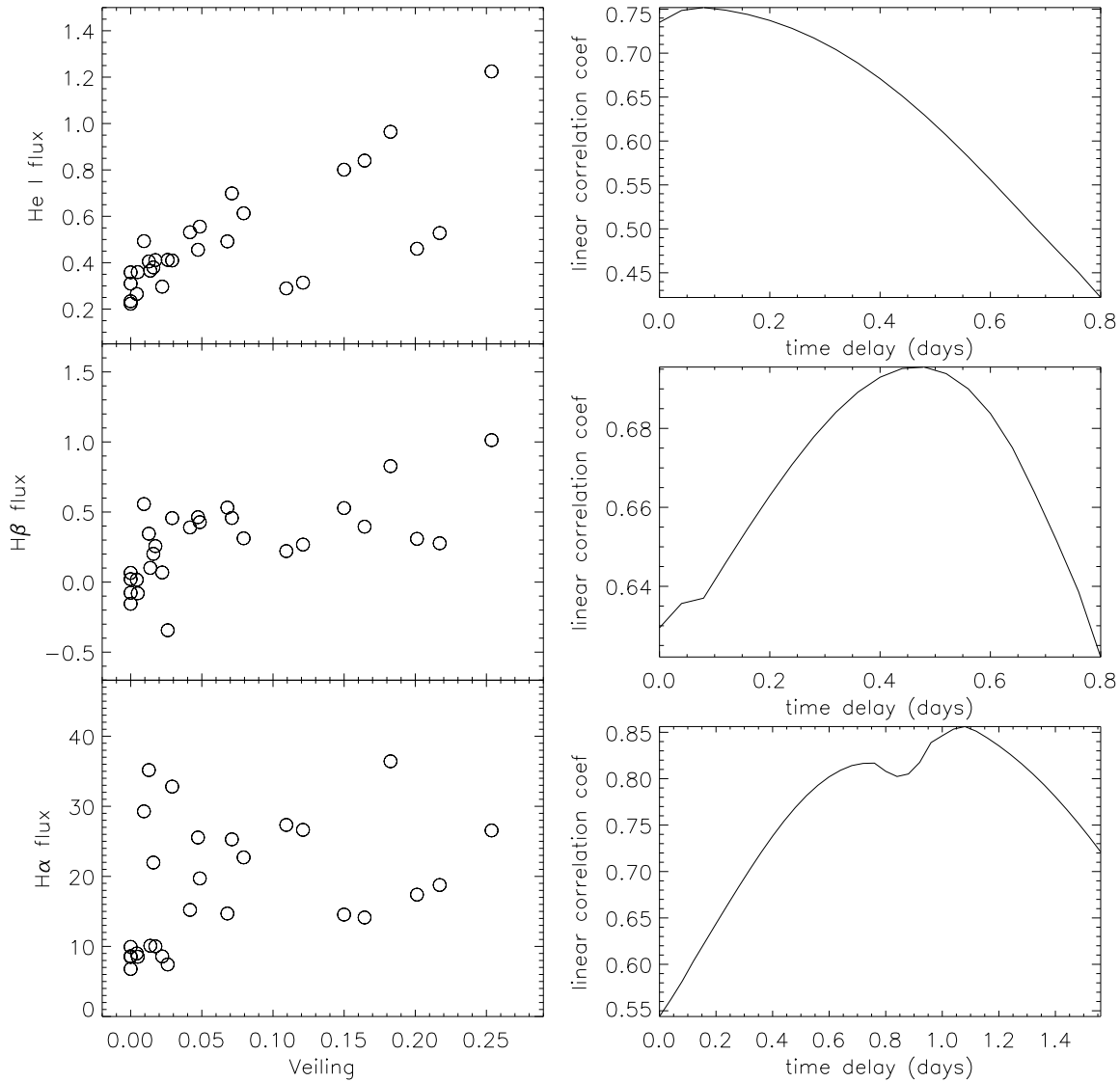}
      \caption{Left panel : The \haln, \hbeta and \heI line fluxes are
        plotted against veiling. Right panel : Time delay between emission
        line flux and veiling variations which provides the best
        correlation between the two quantities (see text). From top to
        bottom: \heI (0.08d), \hbeta (0.44d) and \hal (1.08d). A positive
        time delay means that the line flux varied before the veiling did.
        }
         \label{tdelay_veil}
   \end{figure*}

   We observe a strong correlation of veiling with the HeI line flux (see
   Fig.  \ref{tdelay_veil}). Two pairs of points however significantly
   depart from the veiling-HeI flux correlation at JD=512.5 and 513.5, just
   after the minimum in the first large photometric dip. We come back to
   this in the discussion section.  There is an indication of a weak
   correlation with H$\beta$ line flux but nothing with H$\alpha$. We
   looked for time-delayed correlations between the line fluxes and veiling
   and found that \hal presented a better correlation with veiling if its
   variations occured 1.08 days before the veiling variations (Fig.
   \ref{tdelay_veil}).  The \hbeta results showed the best correlation is
   obtained with a 0.44-day delay (Fig.  \ref{tdelay_veil}) and the \heI
   line is better correlated with the veiling with a time delay of 0.08 day
   (Fig.  \ref{tdelay_veil}). The time lag between line flux and veiling
   variations is larger for lines that are formed farther away from the
   photosphere as predicted by the magnetospheric accretion scenario.

%
%
%
%

%

%
   
\begin{figure}
   \centering
   \includegraphics[width=8cm]{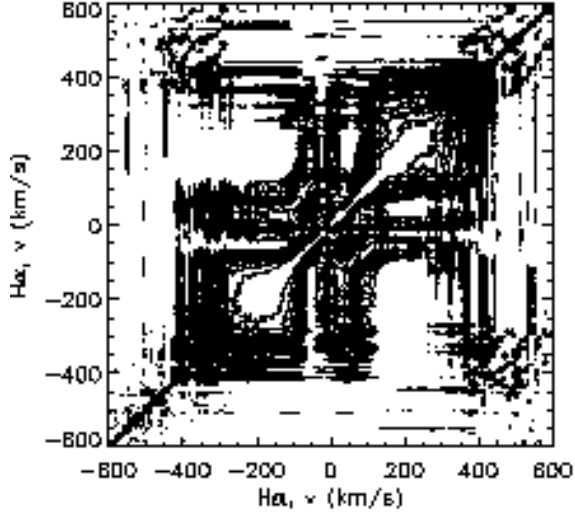}
      \caption{\hal vs. \hal correlation matrix calculated with
      all the observed profiles.
              }
         \label{matrix_all}
   \end{figure}

\begin{figure}
   \centering
   \includegraphics[width=8cm]{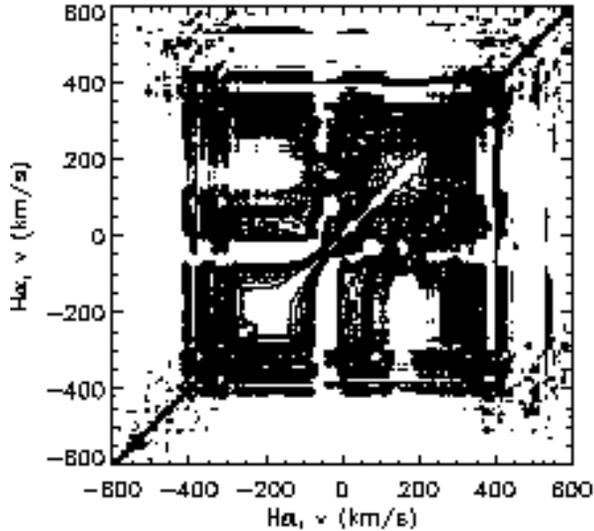}
      \caption{\hal vs. \hal correlation matrix calculated without
        the profiles that presented asymmetric red wings.  }
         \label{matrix_sym}
   \end{figure}

   We calculated correlation matrices for the main emission lines in
   order to investigate how the profile variations are correlated
   across the line. Correlation matrices are 2D contour plots of
   linear correlation coefficients (see Johns \& Basri 1995b). The
   coefficients are calculated, in this work, correlating the time
   variation of each velocity bin of a spectral line with the time
   variation of all the other velocity bins of the same line or of a
   different line.  Using all the observed \hal profiles, the outer
   blue and red wings correlated well with themselves but showed
   almost no correlation with each other (Figure \ref{matrix_all}),
   which is rather unexpected since they are both thought to come from
   the high-velocity regions of the accretion flow.  However, if we
   carefully select only the \hal spectra with symmetric red wings,
   taking away those that showed a lack or an excess of emission
   compared to the blue wing, the outer wings do correlate (Figure
   \ref{matrix_sym}). This indicates that in addition to the low
   velocity absorption in the red wing, there is also something going
   on at high velocities that affects the \hal profile and
   consequently its correlations. Although a redshifted absorption
   component is never clearly seen in the outer red wing of \haln, it
   is probably present and sometimes can be seen in outer red wings of
   \hbeta (see Fig.  \ref{profiles}).  Looking at the symmetric
   profile matrix (Figure \ref{matrix_sym}), the regions that do not
   correlate with the rest of the profile correspond to the blue and
   red low-velocity absorptions. We notice that the redshifted
   absorption at low velocities moves around much more accross the
   profile than the blueshifted one, thus creating a larger region
   with no correlation. No anti-correlated regions were found in the
   matrices, as also noticed Johns \& Basri (1995b) when they
   calculated the \hal correlation matrix of AA Tau.

%

%
   
   \begin{figure}
     \centering \includegraphics[width=8cm]{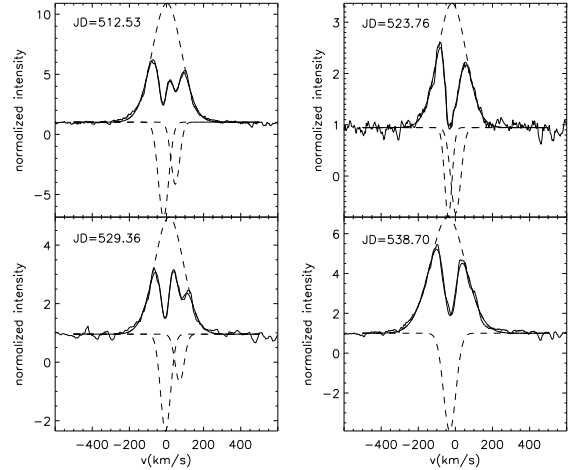}
      \caption{Examples of the decomposition of the \hal line profile using three
        Gaussians. Thin solid line: observed spectrum, dashed lines:
        individual components, thick solid line: all components added
        together.  }
         \label{decomp}
   \end{figure}
%
 
   The \hal and \hbeta lines display more than one component with different
   characteristics and variability. In order to investigate the behavior of
   each line component separately we decomposed the \hal profiles using
   three Gaussians, corresponding to a centered emission, a blueshifted
   absorption and a redshifted absorption (see Fig.~\ref{decomp}). We did
   not make any attempt to fit the high velocity redshifted absorption
   component that is conspicuous only in some \hbeta profiles. The \hal
   decomposition is not always straightforward since there are two
   low-velocity absorption components superimposed on the main emission
   profile. The equivalent width of the absorption components are most of
   the time quite uncertain due to their proximity that makes it hard to
   disentangle them at our present resolution.  The radial velocities of
   the three components however are well determined by the Gaussian
   decomposition.  The radial velocity of the emission component is very
   well constrained by the profile wings and it surprisingly varied from
   $-45$ \kms to $+30$ \kmsn.  The radial velocity of the blueshifted
   absorption component varied from $-38$ \kms to $-5$ \kms and that of the
   redshifted absorption component from $2$ \kms up to $70$ \kmsn.
   
\begin{figure}
   \centering
   \includegraphics{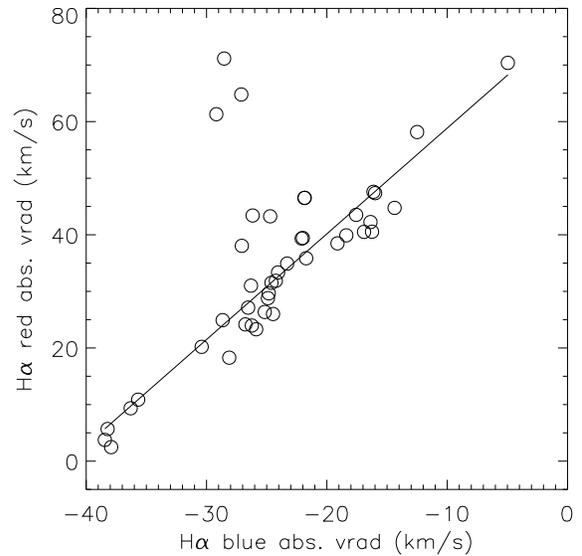}
      \caption{Radial velocities of the \hal red and blue absorption
        components. A tight correlation is observed between the two
        quantities. The three data points that strongly depart from this
        correlation occurred on JD 529.5, 530.5 and 531.5.  }
         \label{ha_redblue}
   \end{figure}

   We found a very good correlation between the radial velocities of the
   \hal blue and red absorption components (Figure \ref{ha_redblue}).
   Except for the three points at the uppermost left side of the plot (JD
   529.5, 530.5, 531.5), the radial velocity of the red absorption
   component changes nearly twice as fast as the blue one, indicating that
   they are not driven by the same processes. The blueshifted absorption
   component is thought to come from a wind, while in general redshifted
   absorption components are related to the accretion process. This result
   is therefore a piece of evidence of a correlation between accretion and
   outflow in AA Tau.
   
   The \hbeta line should have been easier to decompose than \hal but the
   low S/N of our spectra in this region prevents a reliable decomposition
   most of the time.  The \heI line was decomposed with one Gaussian in
   emission and it presented small radial velocity variations, from $-4$
   \kms to $11$ \kmsn (see Fig. \ref{line_flux}).

%

\section{Discussion}

AA Tau's exhibited a more complex pattern of photometric variability
in 1999 than previously observed in 1995. There are qualitative
similarities between the two light curves which suggests the main
source of variability has remained the same between the two
epochs. There are, however, major differences as well. The 1999 light
curve does not appear as periodic as it was in 1995~: the latter light
curve exhibited well defined, large amplitude luminosity dips on a
quasi-periodic timescale of 8.2 days while the new light curve
exhibits two smaller amplitude and asymmetric dips on this timescale
and one of the major luminosity dips disappeared during one cycle. A
rich pattern of variability is also seen in the emission line profiles
and fluxes. Combining the photometric and spectroscopic results we
attempt below to outline a global model for the origin of AA Tau's
variability.

\subsection {Origin of the photometric variations}

The similarity of the light curve recorded for AA Tau in 1999 with
that obtained in 1995 suggests that the same dominant mechanisms were
reponsible for the observed photometric variability at the two
epochs. B99 interpreted the photometric behaviour of this system as
resulting from the obscuration of the stellar photosphere by
circumstellar material orbiting the star at keplerian velocity. The
opacity screen responsible for the eclipses was tentatively identified
with the warped inner edge of AA Tau's circumstellar disk close to the
corotation radius where it interacts with the star's inclined
magnetosphere. This interpretation has subsequently gained support
from the physical modelling of the response of a circumstellar disk to
an inclined dipole as the models predict the development of a non
axisymmetric warp at the inner disk edge (Terquem \& Papaloizou 2000,
Lai 1999).

Arguments against alternative interpretations of AA Tau's light curve were
given in B99. The new light curve and spectroscopic data provide additional
ones that we summarize here briefly.

\subsubsection{A planetary mass companion ?}

The detection of radial velocity variations with a period of 8.3 days and
an amplitude of order 2~\kms (see Fig.~\ref{vrad_phot}) may point to the
existence of a low-mass companion orbiting the star. B99 discussed this
possibility before uncovering the periodic radial velocity variations.
Using the newly derived amplitude of the radial velocity curve, we can
  now derive a maximum mass (assuming a circular orbit) of 20 Jupiter
  masses for the putative companion. However, it is unlikely that such an
  orbiting companion could be responsible for the observed photometric
  variability.  The non-steady character of the photometric light curve
  and, in particular, the temporary disappearance of one of the major
  luminosity dips during one cycle, conflicts with the stability expected
  from the orbital motion of a substellar companion. In addition, the
  radial velocity variations are still present during the photometric
  plateau, i.e., when the large luminosity dip has desappeared (cf.
  Fig.~\ref{line_flux}). Hence, there seems to be no direct connection
  between the source of the large scale photometric variability and the
  source of the radial velocity variations.
  
  Interpreting the radial velocity variations of AA Tau as reflex motion
  induced by an orbiting low mass companion would imply that the orbital
  period of the companion (8.3d) is similar to the stellar rotational
  period (8.2-8.5d). This could possibly result from the companion having
  experienced type II inward migration in the disk which stops at the inner
  disk edge (Lin et al. 1996). In AA Tau, the disk truncation radius lies
  near or at the corotation radius (0.08 AU, see B99), thus leading to an
  orbital period at this radius similar to the stellar rotational period.
  In order to further investigate the possible existence of a substellar
  companion in close orbit around AA Tau, a new spectroscopic monitoring
  campaign is planned for the fall 2003 which will measure the star's
  radial velocity curve over several months.

\subsubsection{Surface spots} 

Alternatively, cold surface spots may be responsible for the periodic
variations of the radial velocity. The 8.3d period of the radial
velocity curve is consistent with the previously reported rotational
period of AA Tau in the range 8.2-8.5d.  We used Petrov et al.'s
(2001) model to compute the radial velocity variations induced by a
dark surface spot.  We find that a 38\degr\ radius spot located at
a latitude of 55\degr\ would produce a periodic modulation of the
stellar radial velocity with an amplitude of 1.6~\kmsn, slightly
smaller than observed. Hence, spots of at least this size would be
required to account for AA Tau's radial velocity amplitude. However,
such a single circular spot would also produce a modulation of the
stellar luminosity with an amplitude of about 0.26 magnitudes in the
V-band. According to the phase of AA Tau's radial velocity curve, the
cold spot would be at maximum visibility around JD 515, 523 and
531. Brightness variations observed at these dates tend to be lower
than those predicted by the model. Hence, we failed to find a spot
model which consistently accounts for AA Tau's radial velocity and
brightness variations.

Independently of the radial velocity variations, ascribing the large
luminosity dips of AA Tau's light curve to photospheric cold spots would
require a projected spot area covering at least 50\% of the visible
photosphere, i.e., a spot radius of about 45\degr\ if circular. For a
stellar rotation period of 8.2d days, it would take at least a couple of
days for such a large spot to go from totally invisible to fully visible,
regardless of its shape. This is inconsistent with the sharp luminosity
decrease observed in the light curve where the system dims by 1 magnitude
on a timescale of a day or less. In addition, huge cold spots observed at
the surface of WTTS are stable on a timescale of several weeks at least and
up to several years (e.g. Petrov et al.  1994).  The missing large
luminosity dip around JD 522 in AA Tau's light curve would conflict with
this expectation. One might assume that AA Tau's rotational period is 16.4d
instead of 8.2d to circumvent this problem, but this is ruled out from
{v$\sin$i}=11.3 $\pm$ +/-0.7 \kms (see above) which, with R$_\star$ = 1.85
$\pm$ 0.15 \rsun\ (B99), yields P = (8.29d +/- 1.2d)$\cdot$sini, i.e. a
maximum rotational period of about 8.5 days for an edge-on system.

Hence, we believe that large luminosity dips cannot be due to cold surface
spots. The smaller, secondary luminosity dips are also unlikely to be due
to cold spots since their duration does not exceed 2 or 3 days, while the
modulation by surface spots would be expected to produce at least 4-day
wide dips for a stellar rotation period of 8.2 days and $\sin$i$\simeq$1.
Furthermore, the small dips exhibit the same color behaviour as the large
ones (cf.  Fig.~\ref{magslop}, Fig.~\ref{cmds}) which suggests a common
cause.

Hot spots are also easily dismissed as a possibly dominant cause of AA
Tau's photometric variability since one would then expect a correlation
between the system brightness and veiling or excess flux due to the
accretion shock. Such a correlation is not observed and the excess flux
actually tends to be larger when the system is fainter (see.
Fig~\ref{ex_veil}).

\subsubsection{Variable circumstellar extinction}

   \begin{figure*}[t]
   \centering
   \includegraphics[width=\textwidth]{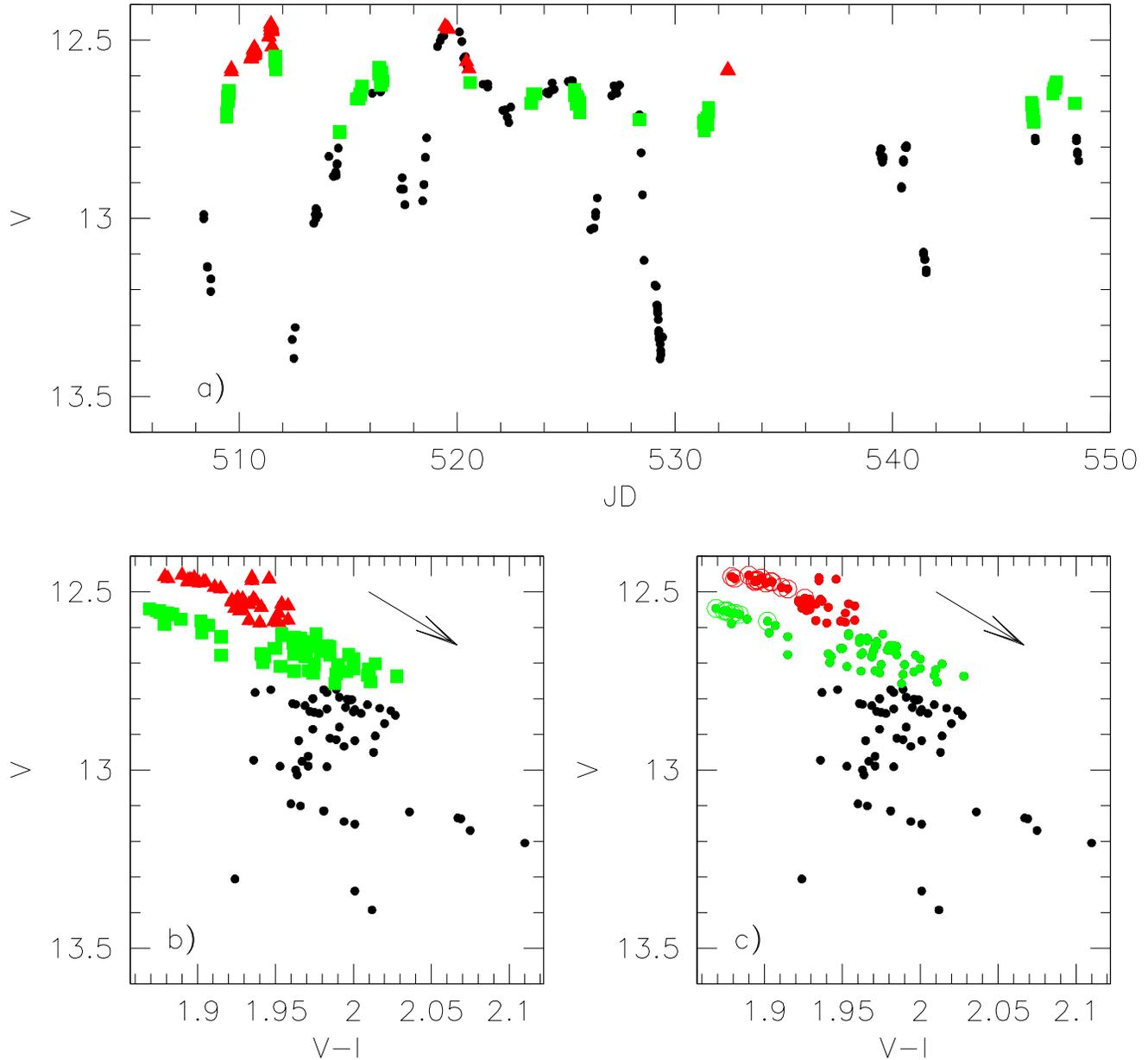}
      \caption{Panel a) AA Tau V-band light curve. At specific dates,
        measurements are represented by large square and triangle symbols
        for reference with panel b). Panel b) V versus V-I diagram: close
        to maximum brightness, the system follows 2 parallel color tracks.
        Symbols are the same as in panel a).  Panel c) V versus V-I
        diagram: the path followed in this diagram by the system on JD 511
        is shown by large empty circles.  As it dimmed, the object suddenly
        flipped from the upper to the lower color track. The arrow in
        panels b) and c) indicates the color slope expected from
        interstellar extinction.}
         \label{branches}
   \end{figure*}
%

Various aspects of AA Tau's photometric variations point to variable
circumstellar extinction: the constant maximum brightness of the light
curve over 150 days (see Fig.~\ref{lc}), the small amplitude of (B-V) color
changes as the system dims except for transient events (see
Fig.~\ref{cmds}), the duration of the dips, their sharp ingress phase and
asymmetric shape, and the lack of a strictly periodic behaviour can all be
accounted for by the partial occultation of the stellar photosphere by
orbiting circumstellar material. Moreover, M\'enard et al. (2003) reported
increased polarisation levels during the faintening episodes, as expected
from the obscuration of the photosphere by circumstellar dust. While other
minor sources of photometric variability might be present as well, neither
stellar spots nor the orbital motion of a companion could account for all
these features simultaneously.
   
Another supporting evidence for circumstellar extinction comes for the
color behaviour of the system. We found above that the (B-V) color is
not significantly affected during luminosity dips (apart from short
timescale events discussed below) while in the (V-R) and (V-I) colors
the system first reddens as it dims. Such a wavelength-dependent
behaviour may result from extinction by dust particules slightly
larger than the interstellar ones, producing opaque occultations at
short wavelengths and interstellar-like reddening at longer ones. The
change of the (V-R) and (V-I) color slopes as the system dims further
(see Fig.~\ref{cmds}) also suggests  non uniform extinction
properties of the occulting screen, which becomes more opaque as the
occulation progresses.

Rapid color changes, best seen in the (B-V) color, are found to occur
at specific photometric phases. Two blueing episodes developped on a
timescale of a few hours around JD 512 and 516, while a rapid
reddening event occurred around JD 529. These events are associated to
the luminosity dips, though they can occur either at the start or in
the middle of the dips. The timescale associated with these events is
comparable to the duration of transit of circumstellar material across
the stellar photosphere (of order of 0.3 days if the occulting
material is located at the corotation radius, cf. B99). The transient
color excursions may thus be related to small-scale ($\leq$ 0.01 AU)
structures in the absorbing material which would indicate that the
occulting screen is somewhat clumpy.
 
A third pattern of color changes is observed on a timescale of a few
hours. Figure~\ref{branches} shows that the (V, V-I) color diagram
exhibits two parallel tracks separated by about 0.1 mag in luminosity
when the system is close to maximum brightness
(Figure~\ref{branches}b). The tracks themselves are roughly parallel
to the reddening slope expected for extinction by small grains. On JD
511.5, at the very start of a large luminosity dip, the system was
first located on the upper track, becoming redder when fainter. It
then suddenly flipped onto the lower track at a slightly bluer color
before starting to redden again as it further dimmed
(Figure~\ref{branches}c). The blueward transition from the upper to
lower (V, V-I) track took less than 3 hours on JD 511.5 and a
qualitatively similar behaviour is observed in the (V, V-R) and (V,
B-V) color diagrams. Other data points on the lower (V, V-I) tracks
are from JD 516.5. This rapid variability is observed close to maximum
brightness, just before the occurrence of photometric dips, and
indicates the sudden appearance of a source of blue continuum excess
flux as the system starts to dim, possibly the accretion shock at the
stellar surface. The strong increase in veiling observed on JD
512.5-513.5 and 516.5-517.5 (see Fig.~\ref{line_flux}), i.e., within a
day after these blueing episodes supports this interpretation.

B99 identified the obscuring material as the warped inner edge of AA
Tau's circumstellar disk close to the corotation radius, as the disk
encounters the stellar magnetosphere. The characteristic timescale of
8.2 days is recovered in the new light curve, thus indicating that the
circumstellar material is still located close to the corotation radius
if in keplerian rotation around the star ($r_{co}\simeq 8.8 R_\star$,
cf. B99). This is also consistent with the short duration of the
ingress phases of the luminosity dips.  However, the structure of the
occulting material appears more complex than it was in 1995. The
asymmetric shape of the major dips suggests a sharp leading edge for
the occulting material and a smoother trailing edge. Also, the light
curve indicates the occurrence of two occulation events per
photometric cycle (cf.  Fig.~\ref{photo_phase}) separated by about
0.4-0.5 in phase, suggestive of the presence of two occulting
structures located at nearly opposite azimuths around the star.
Finally, the depth of the eclipses is shallower in 1999 ($\sim$1 mag)
than it was in 1995 ($\sim$1.5 mag).
   
Because the AA Tau system is suspected to be seen nearly edge-on (see B99),
the occulting material has to lie close to the equatorial plane and is thus
likely associated with the circumstellar dusty disk. A smooth azimuthal
warp of the inner disk, as suggested by B99 to account for the 1995 light
curve, does not produce two occultations per orbital cycle. Terquem \&
Papaloizou (2000) showed that the response of an accretion disk to an
inclined stellar magnetosphere actually produces a warp with two vertical
maxima on the upper side of the inner disk, located at opposite azimuths
and having unequal amplitudes (see their Fig.~6). Such a warp can
qualitatively account for the occurrence of two luminosity dips of unequal
depth within one single orbital period of the inner disk edge. It is also
worth noting that one of the magnetic configurations explored by Terquem \&
Papaloizou (2000) results in a vertical structure for the inner disk warp
which mimics a trailing spiral pattern. Such a warp configuration would
qualitatively account for the asymmetric shape of the main luminosity dips
observed in AA Tau's light curve, with a sharp ingress phase and a slower
return to maximum brightness. In any case, the differences between the '95
and '99 light curves indicate that the structure of the inner disk warp has
changed between the two epochs.

\subsection {Validity of the magnetospheric accretion scenario}

The light curve presented here, as well as that obtained in 1995,
appears to be best interpreted in terms of recurrent occulations of
the central star by circumstellar material located at the warped inner
edge of AA Tau's accretion disk. A likely origin for the inner disk
warp is the response of the disk to an inclined stellar
magnetosphere. B99 proposed a model for AA Tau where the accretion
disk is disrupted by the stellar magnetosphere at a distance of 0.08
AU, resulting in circumstellar material being channelled onto the star
along magnetic accretion columns and eventually hitting the stellar
surface to produce two opposite accretion shocks located at
intermediate latitudes (see their Fig.~10). This model was able to
account for the major characteristics of the luminosity and color
variations of the system in 1995.

   \begin{figure*}
   \centering
   \includegraphics{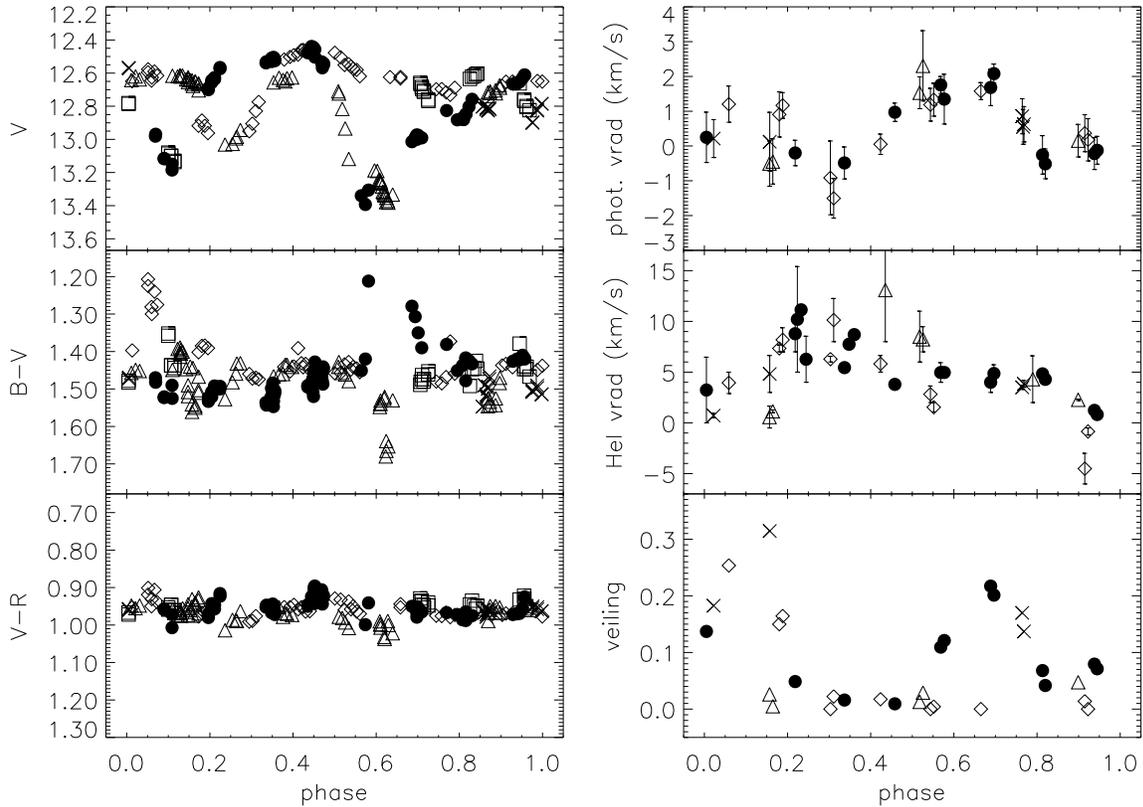}
      \caption{V magnitude, B-V and V-R (left), photospheric radial
      velocity, \heI radial velocity and veiling (right) in phase with
      a 8.2d period (JD$_0$=516). Different symbols represent
      different orbital cycles.  }
         \label{photo_phase}
   \end{figure*}
%
   
   We have now an opportunity to further test this model from the observed
   spectroscopic variations of the system along the photometric cycle.
   Several aspects of the AA Tau spectroscopic analysis seem to confirm the
   general scenario of magnetospheric accretion.
   
   The veiling shows two peaks occuring per 8.2-day cycle and which
   last 3 to 4 days with the highest value around 0.3 (see
   Fig. \ref{photo_phase}). The peaks suggest the presence of two
   rotationally modulated hot spots corresponding to the accretion
   shocks located around the magnetic poles at the stellar
   surface. The B-V color of the system shows some correlation with
   veiling and at least the bluest B-V values clearly correspond to
   the highest veiling values (Fig.  \ref{ex_veil}). This confirms
   that blueing events are associated with veiling variations, both
   being produced by a hot source of continuum flux identified with
   the acretion shocks on the stellar surface.

One of the veiling maxima occur within one day of the center of the
   first deep photometric minima on JD 513 (see
   fig.~\ref{line_flux}), which suggests that the hot spot faces the
   observer at this phase, in agreement with B99's geometrical
   model. Another veiling maximum occurs about 3-4 days later around
   JD 516-517, which would correspond to the maximum visibility of the
   second accretion shock, located on the opposite magnetic pole. Both
   episodes of increased veiling are preceeded by a sudden blueing of
   the system on JD 511.5 and 516.5 (see previous section) which may
   correspond to the appearance of the accretion spots on the stellar
   limb. Note that the veiling maxima seem to have disappeared during
   the photometric plateau (JD~519-525), which indicates that the mass
   accretion rate was much lower during that cycle. The third and
   highest veiling maximum is seen around JD 533 but the sparse
   sampling of both spectral and luminosity variations around this
   date prevent us from associating this event with a clear
   photometric pattern. Figure~\ref{photo_phase} nevertheless
   indicates that this last veiling maximum is in phase with the one
   which occured on JD 516-517.
   
   The veiling also strongly correlates with the \heI line flux
   (Fig.~\ref{tdelay_veil}) which is expected in magnetospheric
   accretion models, since the \heI emission line is thought to form
   at the base of the accretion column close to the accretion
   shock. It is interesting to note that two pairs of HeI flux
   measurements, on JD 512.5 and 513.5, are weaker than expected from
   the overall correlation. Since these dates correspond to the center
   of a large luminosity dip, this indicates that the HeI emission
   region close to the stellar surface is partly occulted at the same
   time as the photosphere. It should be pointed out, however, that
   the apparent veiling is likely overestimated on these dates due to
   the occulation of the photosphere. Since veiling measures the ratio
   between the excess flux and the photospheric flux, the partial
   occultation of the photosphere artificially enhances veiling even
   if the excess flux remains constant. Thus, the 1 magnitude drop in
   the photospheric flux between JD 511 and 512 would result in an
   increase from 0.04 to 0.12 for the veiling. This only partly
   accounts for data points that strongly deviate from the average
   correlation between line flux and veiling. The same effect is seen
   in the correlation between the veiling and the continuum excess
   flux (see Fig.~\ref{ex_veil}). As noted by B99, the sudden increase
   of veiling as the system dims actually reinforces the
   interpretation of AA Tau's photometric variations as being due to
   the occultation of the stellar photosphere.
   
   We see high velocity (150-300 \kmsn) redshifted absorption
   components in the wings of \hbeta and \haln.  These occur most
   clearly at JD=513.5, simultaneously with the veiling peak, and at
   JD=525.5, with no increase of veiling associated to it. Redshifted
   absorptions at high velocity are usually associated to material
   free-falling onto the star along magnetospheric accretion
   columns. They ought to be observed when the hot spot is seen
   through the accretion columns along the line of sight. The
   simultaneous occurrence on JD 513.5 of a high velocity redshifted
   absorptions in the Balmer line profile and of maximum veiling is
   consistent with the accretion column being projected onto the hot
   spot facing the observer. This phase corresponds to the center a
   the large luminosity dip which lasts from JD 511.5 to JD 515.5. The
   other occurrence of a redshifted absorption component in the \hal
   and \hbeta profiles is seen on JD 525.5. Even though veiling is
   weak on this date we argued above that this phase corresponds to
   the maximum visibility of the second accretion shock. The weak
   veiling and line fluxes on this date, located at the end of the
   photometric plateau, are indicative of a strongly reduced mass
   accretion rate onto the star (Fig.~\ref{line_flux}). According to
   magnetospheric accretion models, high velocity redshifted
   absorption components in the Balmer lines can still be seen for
   mass accretion rates as low as 10$^{-9} M_\odot yr^{-1}$ and
   actually become more conspicuous at lower accretion rate as the
   line optical depth decreases (Muzerolle et al. 2001). Weak veiling
   and the appearance of high velocity redshifted absorptions are
   therefore not necessarily contradictory.

   \begin{figure}
   \centering
   \includegraphics[width=0.5\textwidth]{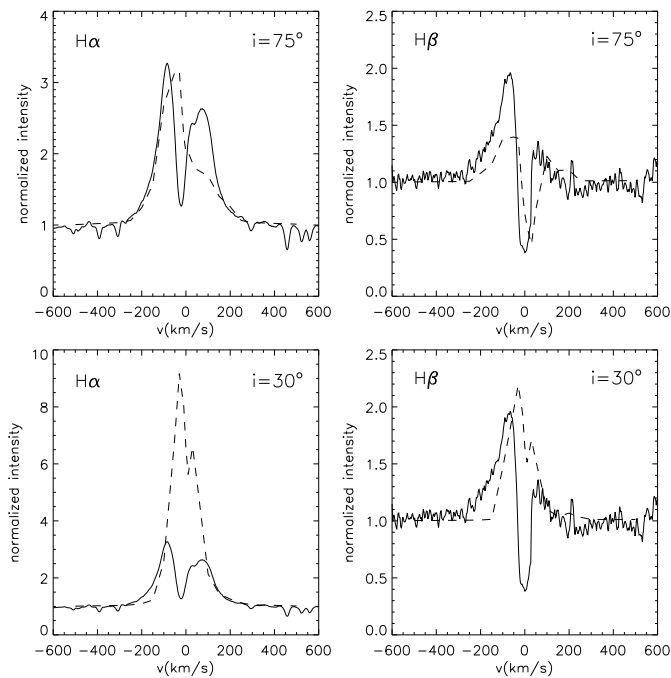}
      \caption{Comparison between the observed \hal (left) and \hbeta
        (right) average profiles (solid lines) and profiles computed from
        magnetospheric accretion models of Muzerolle, Hartmann, \& Calvet
        2001 (dashed lines). The model profiles were computed for two
        inclinations, 75$\degr$ (top) and 30$\degr$ (bottom), a mass accretion
        rate of 10$^{-8} M_\odot yr^{-1}$, a maximum temperature of 7000 K
        in the accretion column and an inner and outer magnetospheric radii
        of 5.2 and 6 R$_\star$, respectively, at the disk plane. Note that
        profiles computed at high inclination provide a much better fit
        than those computed at low inclination.}
         \label{line_models}
   \end{figure}
%
   
   We compared the observed \hal and \hbeta average emission line profiles
   with emission line profiles computed from the magnetospheric models of
   Muzerolle et al. (2001). Figure~\ref{line_models} shows that although
   none of the models perfectly reproduce the observed profiles, the best
   fits are obtained for accretion rates in the range $10^{-8}-10^{-9}$
   \msun ${\rm yr}^{-1}$ and large inclination angles ($i \ge 60\degr$).
   The accretion rate is compatible with measured values for AA Tau,
   $\dot{M}=3.3\times10^{-9}$ and $7.1\times10^{-9}$ \msun ${\rm yr}^{-1}$
   obtained by Gullbring et al. (1998) and Valenti et al. (1993)
   respectively. The low inclination theoretical profiles are much more
   intense and narrower than the observed ones at all phases. The fact that
   only high inclination models fit the observed \hal and \hbeta profiles
   suggests the inclination between the rotation and magnetospheric axis
   cannot be very large. In general, the Balmer line profiles presented a
   redward asymmetry (more emission in the blue than in the red wing) and
   large variability of the red wings, which is overall in agreement with
   the predictions of magnetospheric accretion models.
   
   We noticed a time lag between the \haln, \hbeta and \heI variations and
   the veiling: the lines formed farther away precede those formed close to
   the accretion shock, as expected for a perturbation propagating
   downwards the magnetospheric accretion column. The measured time delays
   are actually quite compatible with the timescale associated with
   free-falling gas in the magnetic funnel. Assuming purely radial motion
   from the disk's truncation radius ($r_m\simeq 8.8 R_\star \simeq 0.08
   AU$, see B99) towards the star, the 1.08 days delay between \hal and
   veiling variations would correspond to gas infall from a distance of
   8.5~R$_\star$ above the stellar surface, which compares well with the
   size derived for AA Tau's magnetospheric cavity. The 0.44 days delay
   measured between \hbeta and veiling would correspond to a radial scale
   of $\sim$6~R$_\star$, while the much shorter 0.08 days delay measured
   for \heI corresponds to about 1~R$_\star$. This indicates, in agreement
   with the predictions of magnetospheric accretion models, that \hal is
   produced in the bulk of the magnetospheric cavity, while \hbeta arises
   from a slightly more compact region and \heI originates close to the
   accretion shock slightly above the stellar surface.
   
   
   We also found a tight correlation between the radial velocities of
   the low-velocity red and blue absorption components in the \hal
   profile (Fig.~\ref{ha_redblue}). This indicates a correlation
   between accretion and ejection signatures, which is predicted by
   the magnetospheric models, since the red absorption is generally
   related to the accretion process and the blue absorption to
   outflows and winds. However, if the wind originated in a region
   spatially associated with the accretion funnel flow, one would
   expect the blueshifted and redshifted velocities to be modulated on
   a rotation timescale and to reach their extreme values
   simultaneously. In contrast, the correlation indicates that the
   largest blueshifted velocitites occur when the redshifted
   velocities are the lowest and conversely. Furthermore, we do not
   find evidence for a rotational modulation of the velocity of the
   absorption components. Instead, Figure~\ref{line_flux} shows a
   monotonic variation over two rotational cycles between JD 508 and
   525. It is therefore quite unlikely that the radial velocity
   variations could result from projection effects modulated along the
   rotational cycle.

   These variations might then reflect intrinsic changes in the
   velocity fields of the inflow ad outflow. A tentative explanation
   for the observed correlation is that the location of the absorbing
   regions moves radially in response to simultaneous changes in the
   inflow and outflow optical depths.  As the accretion rate and wind
   density both increase, the line optical depth becomes larger and
   the absorbing layers move upwards.  This produces a lower
   redshifted absorption velocity since it arises in the upper, lower
   velocity part of the magnetic funnel flow, and a larger blueshifted
   velocity since it forms higher up in the accelerating wind. In this
   scenario, the comoving absorptions components of the H$_\alpha$
   profile would simply reflect the opposite direction of acceleration
   for the inflowing and outflowing material.
   
   This interpretation, however, is not fully supported by the
   observations. It would require that the largest optical depth and
   thus accretion rate occurs around JD 524 when the lowest redshifted
   velocities are observed. In contrast, the line fluxes and veiling
   are the lowest on this date, indicative of a very weak accretion
   rate. Hence, we fail to find a convincing explanation of the
   observed correlation between the velocity of the blueshifted and
   redshifted \hal absorption components in the framework of a steady
   magnetospheric accretion model. We will come back on the origin of
   this correlation in the next section, when we discuss dynamical
   effects associated with the interaction between the inner disk and
   the stellar magnetic field.

%
%
%

\subsection{Evidence for a time-variable configuration: beyond the idealized model }

The spectral and photometric variations of AA Tau, like those of most CTTS,
are obviously more complex than would be expected from a naive axisymmetric
and steady magnetospheric accretion model in which a stable stellar dipole
aligned with the star's rotational axis disrupts the inner part of the
accretion disk. While axial symmetry allows for the modelling of
accretion-ejection structures (e.g. Shu et al. 1994; Ferreira 1997) and for
the computation of line fluxes and line profiles arising in funnel flows
(Muzerolle et al. 2001), the spectrophotometric monitoring of classical T
Tauri stars reveals departures from these models.

Our previous study of the photometric variations of AA Tau already
provided evidence for the accretion flow being channelled along the
lines of a tilted instead of an aligned stellar dipole (B99). Since
then, a surface magnetic field of order of 1-3 kG has been reported
for AA Tau, with no clues however to its topology (Johns-Krull \&
Valenti 2000).  Evidence for a tilted dipole had previously been
reported for the classical T Tauri star SU Aurigae by Johns \& Basri
(1995a). The new photometric and spectroscopic data presented here are
still globally consistent with the magnetospheric accretion scenario
proposed for AA Tau by B99 based only on its photometric variations.
However, some specific aspects of AA Tau's variability are difficult
to account for by assuming a mere steady-state accretion flow along
the lines of an inclined magnetosphere.

The most challenging feature in the observations reported here is the
absence of a main luminosity dip around JD 521-522
(cf. Fig~\ref{replicate}).  If these dips result from the occultation
of the central star by the inner disk warp, as advocated above and in
B99, how can the occulting screen disappear on a timescale of a week
and reappear on the following cycle~? At the time of the missing
occultation, the system also exhibits little spectral variability, the
veiling is minimal as are the line fluxes.  All diagnostics thus
suggest that the system's variability has suddenly shut off for a few
days (JD 519-525, the photometric ``plateau'') and that the accretion
flow onto the star was severely depressed at these dates.

Remarkably enough, the object was quite active on both sides of the
plateau, i.e., during the previous and the following 8.2-day cycles,
exhibiting both large luminosity dips and strong line profile variability.
Moreover, the pattern of variability after the plateau is strinkingly
similar to what it was before, as if the occulting screen had been
suppressed for about the duration of a cycle and had then reformed with
optical and geometrical properties quite similar to those it had before.
Such a repeatability of the variability pattern would be unlikely to happen
if the occultations were produced by independent, free flying dusty blobs
crossing the line of sight as they orbit the star. Instead, it suggests
that the properties of the absorbing material are shaped by an organized
underlying structure.

In B99, we have argued that the obscuring material is to be identified with
the warped inner edge of the accretion disk. The non-axisymmetric warp
itself results from the response of the disk to the inclined stellar dipole
(cf. Terquem \& Papaloizou 2000). The properties of the occulting screen
are thus dictated by the topology of the stellar magnetic field at the disk
inner edge. The stability of the warp configuration, and therefore of the
observed eclipses, ultimately depends on the stability of the magnetic
structure at the disk truncation radius. We thus propose that the ``missing
occultation'' results from the perturbation and subsequent restoration of
the stellar magnetic field at the disk inner edge, leading to the temporary
disappearance of the disk warp and of the associated eclipse as well as to
a severe reduction of the accretion flow onto the star.

The dynamical evolution of a stellar magnetosphere interacting with an
accretion disk has been investigated by several recent numerical models,
with applications to T Tauri stars (e.g. Romanova et al. 2002; Goodson et
al. 1997; Miller \& Stone 1997). These models assume an initial dipolar
configuration and predict that the stellar magnetic field lines threading
the disk expand as they are twisted by differential rotation between the
inner disk and the star. The inflation of stellar field lines also occurs
when accreted material accumulates at the inner disk edge against the
magnetosphere and builds up a pressure gradient that brings the disk
truncation radius closer to the star (Romanova et al.  2002). In most
models, twisted magnetic field lines eventually open, leading to an episode
of strong mass outflow (e.g. Hayashi et al. 1996), and reconnect thus
restoring the initial (dipolar) configuration and the associated accretion
funnel flow onto the star. This evolution is found to repeat itself in a
quasi-periodic manner as originally suggested by Aly \& Kuijpers (1990)
with an associated timescale of order of a few rotation periods (e.g.
Uzdensky et al. 2002a, 2002b; Goodson \& Winglee 1999; Romanova et al.
2002).

We propose that the quiescent phase of AA Tau's variability, which spans
the photometric plateau and associated weak veiling and line fluxes,
corresponds to a phase when the magnetosphere has expanded as described by
these models, thus reducing the accretion flow onto the star as the field
lines become sharply bent at the disk surface (e.g. Romanova et al. 2002).
Models predict that this phase is characterized by a stronger outflow (e.g.
Hayashi et al. 1996; Goodson et al. 1997), conceivably carrying away the
material accumulated at the inner disk edge and thus possibly partly
responsible for the disappearance of the occulting screen. The reduced
accretion rate into the funnel flow may also lower the optical thickness of
the absorbing material at the disk inner edge, thus producing a much
shallower eclipse.  Although we do not have a quantitative measure of the
strength of the wind in the system, we note that the blueshifted absorption
component of the \hal profile reaches below the stellar continuum during
the photometric plateau while it is shallower at all other phases. This may
be an indication that the outflow is indeed stronger when the accretion
flow is depressed. After the inflation phase, field lines reconnect and the
initial magnetic configuration is restored. A new major eclipse is thus
seen around JD 528 as line fluxes start to increase again
(Fig.~\ref{line_flux}). One of the differences between the major eclipses
on JD 511 and JD 528 is that during the latter the system reddens while it
exhibited blueing episodes during the former. This suggests that the
obscuring material was partially optically thin after the restoration of
the initial configuration.


   \begin{figure}
   \centering
   \includegraphics[angle=270,width=0.5\textwidth]{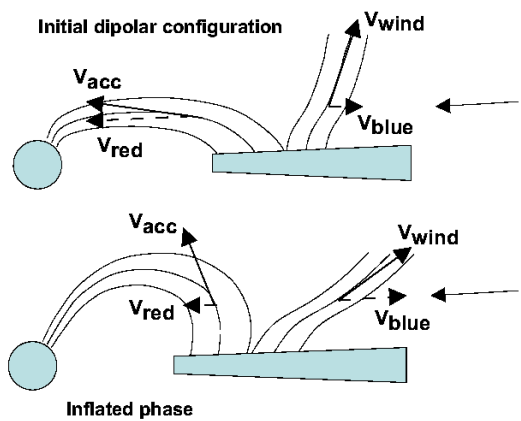}
      \caption{A sketch of the dynamical evolution of the stellar
        magnetosphere interacting with the accretion disk. {\it Top :}
        initial dipolar configuration; {\it bottom :} the magnetosphere has
        expanded as magnetic field lines are twisted by differential
        rotation (see text). The accretion funnel flow onto the star and
        the nearly perpendicular outflow are schematically drawn. The line
        of sight to the system is highly inclined ($i\simeq 75\degr$, shown
        as the arrow on the right). The radial velocities of the inflow and
        outflow are shown. In the initial configuration, the velocity field
        of the absorbing layers in the accretion column is nearly parallel
        to the line of sight, thus yielding the highest radial velocity of
        the redshifted \hal absorption component. In the expanded
        magnetosphere configuration, the same absorbing layers make a large
        angle with the line of sight, thus resulting in small observed
        radial velocities. The outflow being nearly perpendicular to the
        inflow, the opposite behaviour is observed for the radial velocity
        of the blueshifted absorption component in the \hal line profile.}
         \label{inflation}
   \end{figure}
   
   This dynamical scenario of the interaction between the inner disk and
   the stellar magnetosphere may additionally offer an explanation to the
   observed correlation between the radial velocities of the blueshifted
   and redshifted absorption components of the \hal profile
   (Fig.~\ref{ha_redblue}). The qualitative interpretation we propose for
   this correlation is schematically depicted in Figure~\ref{inflation}. In
   the initial configuration, the \hal absorbing layers located in the
   upper part of the accretion funnel flow are nearly parallel to the line
   of sight, thus yielding the highest projected redshifted velocities. In
   the inflated phase, the field lines have strongly expanded and the same
   layers are now at a large angle to the line of sight, thus yielding the
   lowest projected redshifted velocities. In most accretion-ejection
   models, the outflow is nearly perpendicular to the inflow in the
   vicinity of the disk-magnetosphere boundary (e.g. Shu et al. 1994;
   Ferreira 1997). Then, the lowest projected blueshifted velocities are
   expected to occur in the initial configuration, when the outflow is
   nearly perpendicular to the line of sight, while in the inflated phase
   the outflow is bent towards the observer (assuming the angle between the
   inflow and the base of the outflow has not changed significantly) and
   thus yields the highest projected blueshifted velocities. Slowly
   evolving projection effects resulting from the inflation of the
   magnetosphere would thus account for the simultaneous radial velocity
   variations of the blue and red absorption components in the \hal
   profiles. This would provide a natural explanation for the fact that the
   highest redshifted velocities are observed when the blueshifted
   velocities are the lowest and conversely, an observational result which
   had no straightforward explanation in the framework of a static
   magnetospheric accretion model.
   
   If this interpretation is correct, {\it the radial velocity of the
     redshifted \hal absorption component yields a direct measurement of
     the expansion of the magnetic field lines} close to the
   disk-magnetosphere boundary, the velocity decreasing as the magnetic
   field line inflates. We can thus use this diagnostics to trace the
   evolution of the magnetospheric structure as it interacts with the disk.
   The radial velocity curve of the redshifted absorption component is
   shown in Figure~\ref{line_flux}. As noted above, the timescale for its
   radial velocity variations is longer than the rotation period of the
   system. Fig.~\ref{line_flux} shows that the radial velocity first
   steadily decreases from the start of the observations on JD 506 down to
   its lowest value reached on JD 525, which would indicate that the
   magnetosphere expands over at least this time span. From JD 525 on, the
   radial velocity rapidly increases to reach its maximum value on JD 530
   and then starts to decrease again. This phase would then correspond to
   the restoration of the initial magnetic configuration over JD 525-530
   and the beginning of a new inflation cycle starting on JD 530.
   
   The dynamical evolution of the magnetosphere deduced from the projected
   velocity of the redshifted \hal absorption component {\it alone} is
   globally consistent with the spectro-photometric evolution of the system
   : the phase of quiescent activity associated to a reduced accretion rate
   onto the star (photometric plateau, lowest veiling, weakest line fluxes,
   smallest line variability) is observed from JD 518 to 525, i.e., at the
   end of the expansion phase (JD $\leq$506-525). The reapparition of the
   eclipses, the increase of line fluxes and of veiling all happen between
   JD 526 and JD 532, as the initial configuration is restored (JD
   525-530). Note that three measurements strongly depart from the
   correlation shown in Fig.~\ref{ha_redblue}. They occur on JD 529.5,
   530.5, and 531.5 precisely at the time the initial configuration is
   re-established. On JD 529.5, large variations of the radial velocity of
   the blueshifted component are seen to occur on a timescale of a few
   hours (see Fig.~\ref{line_flux}). This suggests that, superimposed onto
   the slowly evolving magnetospheric structure, transient ejection events
   occur as a new magnetospheric cycle starts.
   
   Thus, independent observational diagnostics of the magnetospheric
   accretion process can be consistently accounted for by a dynamical
   description of the interaction between the disk and the stellar
   magnetosphere. This result provides strong support to the recent
   numerical simulations which predict a time variable behaviour of the
   disk-magnetosphere interface and may have important implications for the
   origin of the spectro-photometric variability of classical T Tauri
   stars, for the regulation of their angular momentum, and for the origin
   and short-term variability of outflows in young objects.

\section{Conclusions}

We have shown that the photometric and spectroscopic variations of AA
Tau on days to weeks timescales are globally consistent with the
concept of magnetospheric accretion in this system. Its inner
accretion disk is truncated at a distance of about 0.1 AU from the
stellar surface by the strong stellar magnetosphere. The large scale
magnetospheric structure is inclined relative to the disk plane which
leads to the developement of a non axisymmetric warp at the inner disk
edge. The warp corotates with the star and is responsible for the
occurrence of the eclipses observed in AA Tau's light curve. As the
eclipses proceed, high velocity inverse P Cygni profile are
episodically observed at \hbeta and veiling is maximum, which suggest
accretion columns are seen against the hot accretion shock at this
phase. The size of the magnetospheric cavity, about 8R$_\star$ as deduced
by assuming keplerian rotation at the inner disk edge, is consistent
with the time delay we measured between lines and veiling variations
as non steady accretion propagates downwards the accretion column,
from the inner disk edge to the stellar surface on a timescale of one
day.

The global structure of the magnetospheric accretion region appears to vary
on a timescale of order of a month. We observed the sudden disappearance of
eclipses together with a strong reduction of line fluxes and veiling for a
few days.  During this episode, the inner disk warp had apparently vanished
and the accretion flow onto the star was severely depressed which suggests
that the magnetic configuration at the inner disk edge has been disrupted.
A few days later, the initial magnetic configuration was restored as
indicated by eclipses resuming with a similar depth and shape as prior to
the disruption event and line fluxes and veiling increasing again. The
phase of reduced activity may correspond to a state where the field lines
had strongly expanded and perhaps opened under the action of differential
rotation between the inner disk and the star. In support to this
interpretation, we find a smooth variation of the radial velocity of
accretion and outflow diagnostics in the \hal profile which is best
interpreted as reflecting the slowly inflating magnetosphere on a timescale
of a month, followed by its disruption and the restoration of the initial
magnetic configuration. Recent numerical simulations describe such a
magnetospheric inflation process and predict they are cyclic. The results
reported here may constitute the first clear evidence for the existence of
magnetospheric inflation processes occurring in CTTS on a timescale of a
month, though additional observations will be needed to assess their cyclic
nature.

The observed variability of AA Tau is thus a complex mixture of rotational
modulation by hot spots, variable circumstellar extinction, non steady
accretion and varying mass loss, all of which appear to be consistent with
a dynamical view of the interaction between the inner disk and an inclined
large-scale stellar magnetosphere. The magnetospheric accretion process
appears to be time dependent on all scales, from hours for non steady
accretion to weeks for rotational modulation and months for global
instabilites of the magnetospheric structure. AA Tau exhibits clear
signatures of these various processes mostly because is it seen at a high
inclination which maximizes the amplitude of variability. However, since AA
Tau has otherwise properties of a very typical classical T Tauri star it is
very likely that the processes observed in AA Tau are also instrumental in
other CTTS, though more difficult to diagnose when the systems are seen at
a lower inclination (e.g. Chelli et al. 1999; DeWarf et al. 2003).

The highly dynamical and time dependent nature of the magnetospheric
accretion process may have implications which remain to be explored for a
number of issues. For instance, the transfer of angular momentum between
the star and the disk (and the wind) is certainly quite complex and time
variable (Agapitou \& Papaloizou 2000). Whether this variability affects
the angular momentum evolution of CTTS on the long term is however unclear,
since it takes about 10$^5$yr to the star to react to angular momentum
gains or losses which will presumably smooth out the effects of short term
variability. The dynamical nature of the magnetospheric accretion process
may also bring clues to the origin of the short-term variability of CTTS
outflows (Woitas et al. 2002, Lopez-Martin et al. 2003). It would be
interesting to search for wind or jet variability on a timescale of a
month, as could be expected from a cyclical evolution of the magnetospheric
structure.  One of the important implications of the results reported here
is related to the origin of CTTS near infrared excess, which is often used
as a quantitative diagnostics of mass accretion in the disk and usually
modelled in the framework of standard $\alpha$-disk models. AA Tau's result
suggest that the structure of the inner disk is in fact strongly modified
from its interaction with the stellar magnetosphere.  The development of a
disk warp can be expected to considerably increase the illuminated fraction
of the inner disk, possibly leading to high veiling values in the near-IR
as measured by Folha \& Emerson (1999) and Johns-Krull \& Valenti (2003).
Also, large scale instabilities of the magnetospheric structure at the
inner disk edge can be expected to produce rapid and large near infrared
variability, as observed in some systems (Carpenter et al. 2001; Eiroa et
al. 2002).

Finally, most MHD models of magnetospheric accretion developped so far
assume axisymmetry and steady-state inflows/outflows. These assumptions are
clearly an oversimplification of the highly dynamical processes at work in
the interaction between the inner disk and the stellar magnetic field. In
view of these limitations, it may not be surprising that some of the basic
predictions of these models are not always confirmed by snapshot
observations which catch the systems at a particular phase of their short
term evolution (e.g. Johns-Krull \& Gafford 2002). Whether these models
remain valid to describe the evolution of these systems on the long term,
once the short term variability reported here has been smoothed out over
much longer timescale, remains to be seen.




\begin{acknowledgements}
  
  We thank J. Ferreira for discussions on MHD accretion-ejection models, M.
  Mouchet for providing us with bibliography on possibly related processes
  at work in cataclysmic variables, and J. Steinacker for discussing
  radiative transfer issues in the framework of the magnetospheric
  accretion model. This work was supported by a NATO Science Program grant
  (PST.CLG.976194). S.H.P.A. acknowledges support from CNRS and CAPES
  (PRODOC program). M.F. is partially supported by the Spanish grant
  AYA2001-1696.

\end{acknowledgements}

\end{document}